\documentclass{article}

\usepackage{microtype}
\usepackage{graphicx}
\usepackage{subcaption}
\usepackage{booktabs}
\usepackage{tikz}
\usepackage{pgfplots}
\usepackage[most]{tcolorbox}
\pgfplotsset{compat=1.18}
\usepackage{multirow}
\usepackage{hyperref}

\usepackage{amsmath}
\usepackage{amssymb}
\usepackage{mathtools}
\usepackage{amsthm}
\usepackage[dvipsnames,svgnames,x11names,table]{xcolor}

\usepackage{tikz}       % 绘图支持
\usepackage{array}      % 表格列格式
\usepackage[table]{xcolor}

\definecolor{TopOneBlue}{RGB}{1,178,176}     % 深蓝
\definecolor{TopTwoBlue}{RGB}{161,226,225}    % 中蓝
\definecolor{TopThreeBlue}{RGB}{180,232,231} % 浅蓝

\definecolor{BotOneRed}{RGB}{186,97,102}      % 深红
\definecolor{BotTwoRed}{RGB}{213,160,163}      % 中红
\definecolor{BotThreeRed}{RGB}{236,213,214}  % 浅红
\newcommand{\best}[1]{\cellcolor{TopOneBlue}\textcolor{white}{\textbf{#1}}}
\newcommand{\secondbest}[1]{\cellcolor{TopTwoBlue}\textcolor{white}{#1}}
\newcommand{\thirdbest}[1]{\cellcolor{TopThreeBlue}{#1}}

\newcommand{\worst}[1]{\cellcolor{BotOneRed}\textcolor{white}{#1}}
\newcommand{\secondworst}[1]{\cellcolor{BotTwoRed}\textcolor{white}{#1}}
\newcommand{\thirdworst}[1]{\cellcolor{BotThreeRed}{#1}}

\definecolor{headerbg}{RGB}{249,247,247} % 很浅的冷灰（可按需调整）

\usepackage{tcolorbox}
\usepackage[preprint]{icml2026}

\usepackage{amsmath}

\usepackage[capitalize,noabbrev]{cleveref}

\theoremstyle{plain}

\theoremstyle{definition}

\theoremstyle{remark}

\usepackage[textsize=tiny]{todonotes}

\icmltitlerunning{AutoMIA: Improved Baselines for Membership Inference Attack via Agentic Self-Exploration}

\begin{document}

\twocolumn[
\icmltitle{AutoMIA: Improved Baselines for Membership Inference Attack via Agentic Self-Exploration}

\icmlsetsymbol{equal}{$\ddagger$}
\icmlsetsymbol{cor}{$\dagger$}

\begin{icmlauthorlist}
\icmlauthor{Ruhao Liu}{equal}
\icmlauthor{Weiqi Huang}{equal}
\icmlauthor{Qi Li}{equal}
\icmlauthor{Xinchao Wang}{cor}
\end{icmlauthorlist}

\icmlcorrespondingauthor{Xinchao Wang}{xinchao@nus.edu.sg}

\printAffiliationsAndNotice{}
\begin{center}
{\small $\ddagger$Equal contribution \, $\dagger$Corresponding Author}
\end{center}
\vspace{-0.4em}
\begin{center}
{\small National University of Singapore}
\end{center}
\vspace{-0.4em}
\begin{center}
\url{https://github.com/amiya-special/AutoMIA}
\end{center}

\vskip 0.1in
]

\begin{abstract}
Membership Inference Attacks (MIAs) serve as a fundamental auditing tool for evaluating training data leakage in machine learning models. However, existing methodologies predominantly rely on static, handcrafted heuristics that lack adaptability, often leading to suboptimal performance when transferred across different large models. In this work, we propose \texttt{AutoMIA}, an agentic framework that reformulates membership inference as an automated process of self-exploration and strategy evolution. Given high-level scenario specifications, AutoMIA self-explores the attack space by generating executable logits-level strategies and progressively refining them through closed-loop evaluation feedback. By decoupling abstract strategy reasoning from low-level execution, our framework enables a systematic, model-agnostic traversal of the attack search space. Extensive experiments demonstrate that AutoMIA consistently matches or outperforms state-of-the-art baselines while eliminating the need for manual feature engineering.

\end{abstract}

\begin{figure*}[t]
    \centering
    \resizebox{\textwidth}{!}{%
    \begin{subfigure}[t]{0.45\textwidth}
        \centering
        \includegraphics[width=\linewidth]{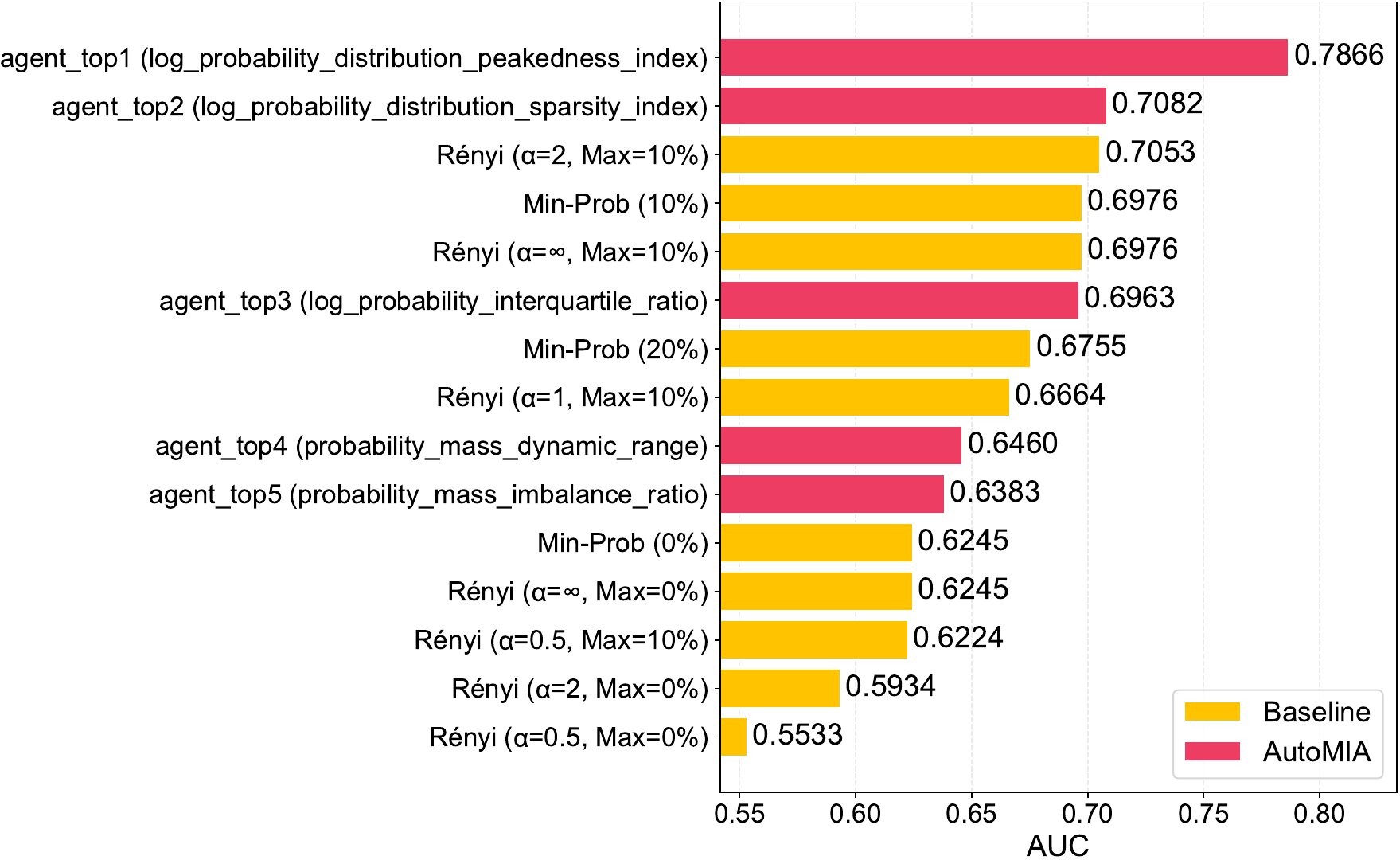}
        \label{fig:a}
    \end{subfigure}
    \hfill
    \begin{subfigure}[t]{0.28\textwidth}
        \centering
        \includegraphics[width=\linewidth]{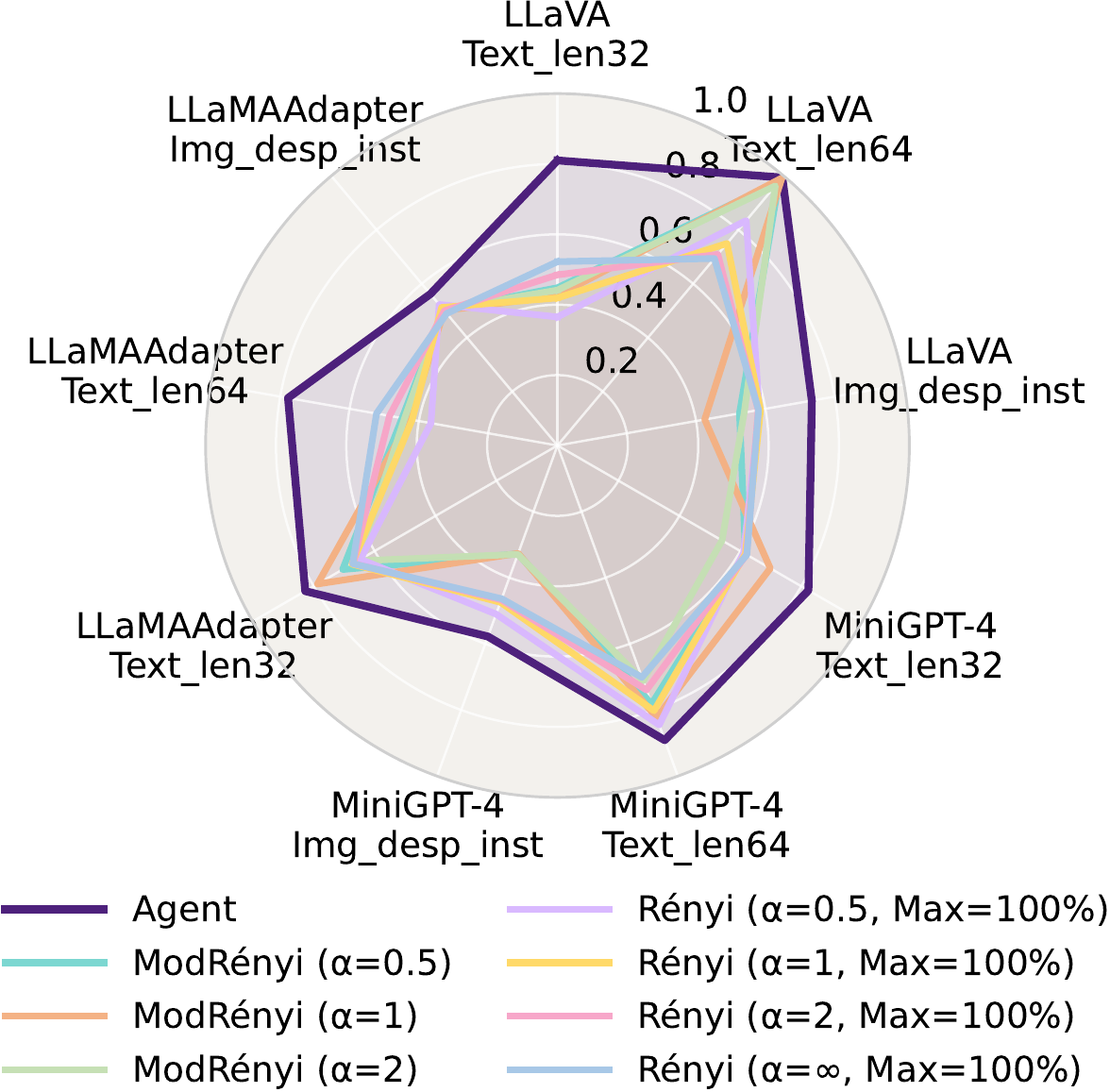}
        \label{fig:b}
    \end{subfigure}
    \hfill
    \begin{subfigure}[t]{0.23\textwidth}
        \centering
        \includegraphics[width=\linewidth]{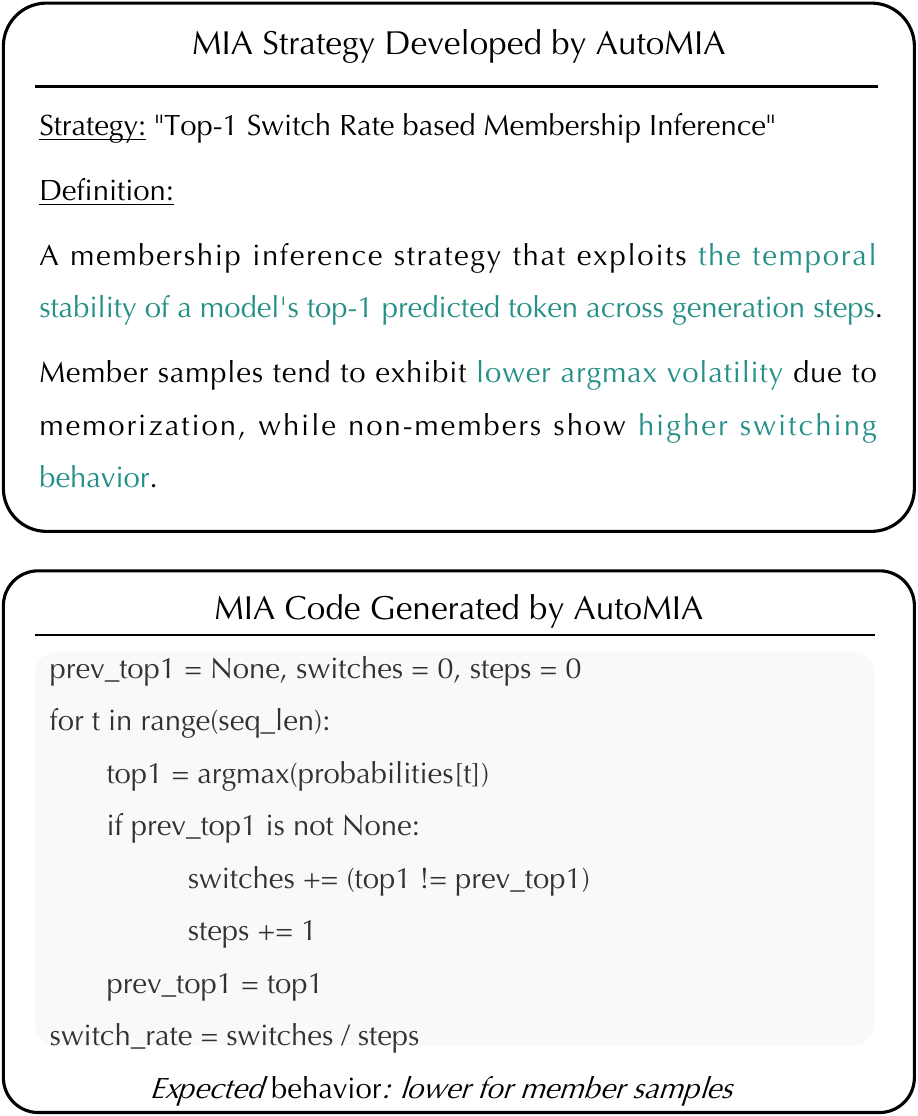}
        \label{fig:c}
    \end{subfigure}
    }
    \vspace{-4mm}
    \caption{
    Performance comparison between AutoMIA and baselines. \textbf{Left:} Comparison of the top five AutoMIA-discovered metrics and the top ten handcrafted baselines on the DALL·E dataset with LLaVA as the victim model. \textbf{Middle:} Comparing text-only membership inference performance across three target models (LLaVA, MiniGPT-4, and LLaMA-Adapter) under multiple dataset settings. \textbf{Right:} An example of an AutoMIA-generated attack strategy, showing its high-level definition alongside the corresponding executable code.
    }
    \label{fig:step_simi}
    \vspace{-2mm}
\end{figure*}

\section{Introduction}

The widespread deployment of large foundation models~\cite{yang2025qwen3,li2024llava,zhang2026makegeometrymatterspatial,feng2025can,feng2025rewardmap} has intensified concerns regarding data privacy~\cite{carlini2021extracting,wang2025towards,li2024data,li2025every,liang2022accmyrinx,liang2022escalated,yin2026refinement,song2025idprotector,song2024anti,ci2024ringid}. Membership Inference Attacks (MIAs)~\cite{shokri2017membership} serve as a fundamental tool in this domain, aiming to determine whether a specific sample was used during training. Successful MIAs can expose sensitive information, making them a standard tool for evaluating privacy leakage~\cite{hu2022membership}.

Existing MIAs typically rely on handcrafted strategies exploiting statistical discrepancies like confidence or entropy~\cite{salem2018ml,yeom2018privacy}. While effective in isolated scenarios, these static heuristics are often tightly coupled to specific tasks and require expert feature engineering~\cite{Carlini2021MembershipIA,Li2024MembershipIA}. Critically, prior work lacks a unified mechanism for strategy exploration; attack design is treated as a manual, isolated stage, limiting scalability and the discovery of effective strategies for different large models. Consequently, designing new attacks becomes highly labor-intensive.

Recent advances in agentic reasoning~\cite{yao2022react,xi2025rise,li2026sponge} motivate a key question: \textit{Can we reformulate membership inference strategy discovery as an automated procedure?}
Building on the success of existing attack strategies, such a reformulation has the potential to further improve attack effectiveness while avoiding extensive manual design and intervention. Despite growing interest in automated safety analysis~\cite{Deng2023MASTERKEYAJ,Chao2023JailbreakingBB,yu2025discrete,xiong2026anatomy}, extending such automation to membership inference is far from straightforward. Unlike prompt-level jailbreaks that yield immediate feedback~\cite{Mehrotra2023TreeOA,liu2024autodan}, MIAs operate on noisy, distribution-level signals without explicit refusal boundaries. This makes automated refinement challenging, as the agent must handle subtle statistical shifts rather than overt safety violations.

In this work, we propose \texttt{AutoMIA}, \textbf{the first} framework for \textbf{automatically discovering membership inference strategies across large language and multimodal models}, addressing these challenges through closed-loop self-exploration. To overcome the difficulty of learning from noisy statistical signals, AutoMIA does not optimize for single-query success; instead, it iteratively generates executable logits-level code and refines it based on aggregated feedback (e.g., AUC scores) from dataset-level evaluations. To address credit assignment without explicit refusal boundaries, we use AutoMIA with a history-aware reasoning process: within a sliding context window, it contrasts high-performing strategies with weaker ones to distill effective attack logic and iteratively refine it into stronger strategies. This design enables systematic exploration of the attack space while being query-efficient and robust to noisy, non-differentiable feedback. Extensive experiments on different datasets and models consistently indicate that existing methods leave significant room for further improvement; for example, as shown in Fig.~\ref{fig:step_simi}, AutoMIA substantially outperforms baselines across multiple evaluation tasks, achieving both higher success rates and broad applicability.

\section{Related Work}

\textbf{Membership Inference Attacks.} 
Membership inference attacks (MIAs) aim to determine training set inclusion, representing a fundamental privacy, it has been studied under different access assumptions, including white-box, black-box, and grey-box settings~\cite{nasr2019comprehensive,salem2018ml,carlini2021extracting,li2025vid}. Most MIAs fall into two categories: metric-based attacks utilizing handcrafted statistics like confidence, entropy, or Min-K\%~\cite{song2019membership,shi2023detecting,zhang2024min}, and shadow model--based attacks that approximate the target model's behavior via surrogates~\cite{shokri2017membership}. While effective in specific scenarios, both paradigms rely heavily on manual strategy design and often exhibit limited adaptability across heterogeneous models. Recent work extends MIAs to large language models, multimodal models, and retrieval-augmented systems, revealing new privacy leakage channels but largely retaining handcrafted attack pipelines~\cite{wen2024membership,li2024membership,wang2025rag}. These limitations motivate the need for more automated and adaptive MIA frameworks.

\textbf{LLM-Based Agents and Safety.} 
Large language model--based agents enable autonomous planning and multi-step reasoning to execute complex workflows~\cite{xi2025rise,li2026sponge}. These capabilities have been extensively explored in security analysis, both as sources of new vulnerabilities (e.g., tool misuse~\cite{wang2025shadows}) and as active instruments for defensive evaluation. In the latter context, systems like AttackPilot~\cite{wu2025attackpilot} and IAAgent~\cite{wuiaagent} demonstrate that agents can autonomously conduct inference attacks by iteratively refining queries, while other works explore agent-based privacy red-teaming to induce training data leakage~\cite{nie2024privagent} or target retrieval-augmented architectures~\cite{wang2025rag}. However, unlike prior agent-based attacks that typically focus on specific pipelines, our work formulates membership inference as a unified, agent-driven process with explicit strategy generation and feedback-based refinement under grey-box constraints.

\section{Problem Setting and Challenges}
\label{sec:problem_setting}

\noindent\textbf{Notation. }
Let $\mathcal{V}$ denote the vocabulary set. An input sample is denoted as $x = (I, X_{\text{ins}})$, where $I$ represents the image input and $X_{\text{ins}}$ represents the textual instruction context. In this work, we focus on a target Vision-Language Model (VLM), denoted as $M$. The model accepts the multimodal input $x$ and produces logits-level features, denoted as $\mathbf{o}$. We use $\mathcal{D}_{\text{train}}$ to represent the target dataset containing the multimodal samples used during the model's training process.

\noindent\textbf{Adversary's Goal. }
We follow the standard definition of Membership Inference Attacks (MIAs) as described in~\cite{shokri2017membership}. Given a target VLM $M$, the adversary aims to determine whether a specific sample $x$ was used during the training stage of $M$. We formulate this attack as a binary classification problem managed by an attack strategy (implemented as executable code $\mathbf{p}$). The strategy takes the model's logits output $\mathbf{o}$ as input and computes an inference score $S = \mathbf{p}(\mathbf{o})$. The membership detector $\mathcal{A}(x; M)$ makes its decision by comparing this score with a threshold $\tau$:
\begin{equation}
\mathcal{A}(x; M) = \mathbb{I}(\mathbf{p}(\mathbf{o}) > \tau),
\label{eq:membership_score}
\end{equation}
where $\mathbb{I}(\cdot)$ is the indicator function that outputs 1 (member) if the condition holds, and 0 (non-member) otherwise.

\noindent\textbf{Adversary's Knowledge. }
Following the standard MIA setup~\cite{li2024membership}, we assume a \textbf{grey-box} scenario where the adversary can query the target model using the image and instruction context, and is allowed to access the tokenizer, output logits $\mathbf{o}$, and generated text. However, the adversary has no knowledge of the training algorithm, gradients, or the specific parameters of the target model.

\begin{figure}[t]
    \centering
    \includegraphics[width=\columnwidth]{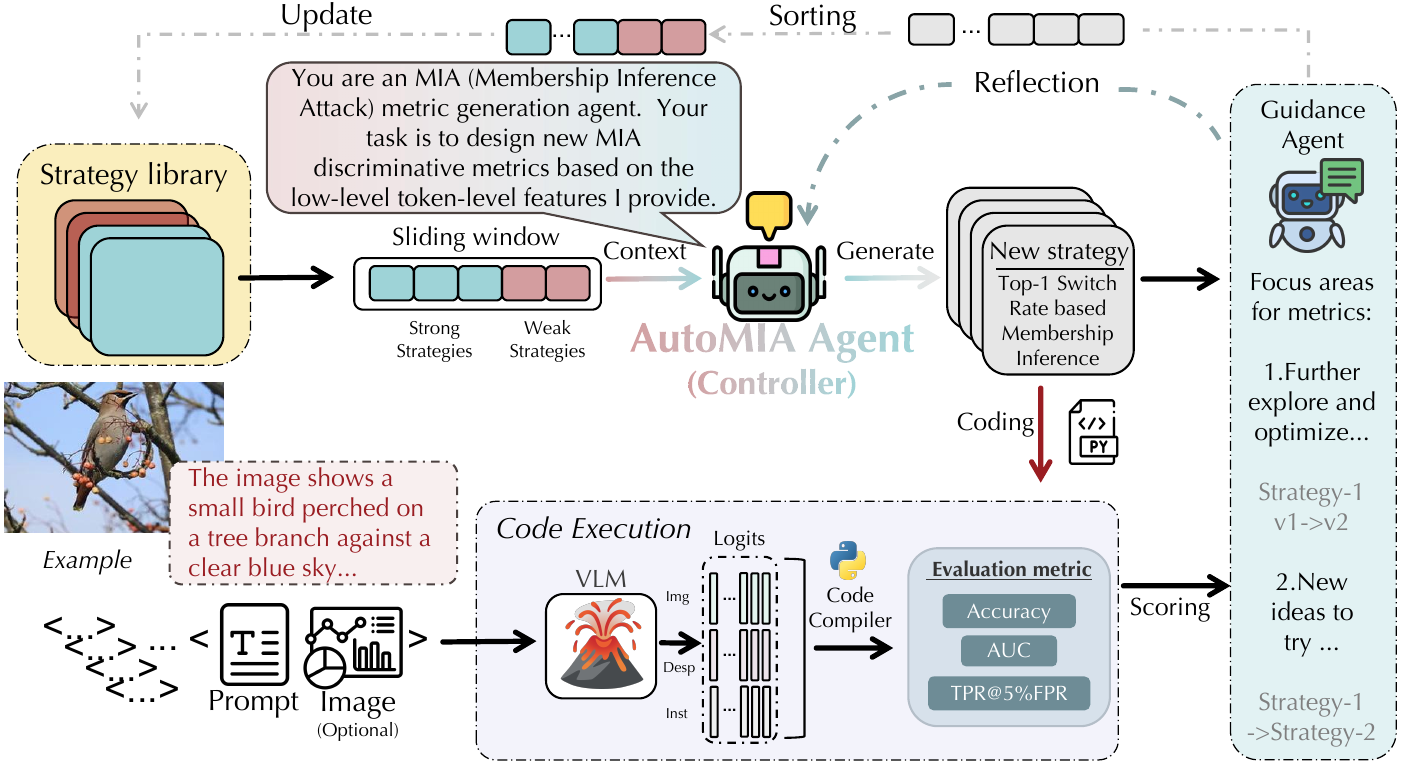}
    \caption{
        Overview of the AutoMIA framework. 
        The system operates as a closed loop where the AutoMIA agent generates strategies based on historical context, the \textit{Code Execution} module runs attacks against target VLMs, and the Guidance agent provides evaluation feedback to refine the \textit{Strategy Library}.
    }
    \label{fig:automia}
    \vspace{-6mm}
\end{figure}

\noindent\textbf{Why not Black-box? }Although the majority of prior MIA studies focus on the grey-box setting~\cite{shokri2017membership,carlini2021extracting,Carlini2021MembershipIA,li2025vid, mattern2023membership,li2024membership,liu2022membership,hu2022membership,Li2024MembershipIA}, black-box attacks remain an important and widely discussed threat model. In this work, we deliberately focus on the grey-box setting, not as a weaker alternative, but as a means to explore the upper bound of membership inference attacks under favorable access conditions. From a practical perspective, the grey-box setting is also well aligned with internal auditing and privacy risk assessment scenarios. In many real-world deployments, training data are not publicly disclosed, while model owners or auditors have full access to model parameters and intermediate outputs. In such cases, privacy evaluation naturally takes place in a grey-box or white-box regime rather than a strictly black-box one. Moreover, the victim models and target datasets used in our experiments are well-designed benchmarks adopted by prior work, serving as controlled testbeds to evaluate attack effectiveness. While these datasets do not aim to fully replicate real-world deployment conditions, they allow us to systematically study attack behavior and isolate the contribution of automated agentic exploration. 

\noindent\textbf{Challenges.}
Reformulating membership inference as an automated agentic process introduces distinct difficulties compared to traditional handcrafted approaches or other automated safety evaluations (e.g., jailbreaking~\cite{liu2023autodan,liu2024autodan}): 

\textbf{(i) Distribution-Level Signals and Absence of Explicit Boundaries.} Unlike prompt-level jailbreak attacks that yield immediate binary success signals (e.g., a harmful response)~\cite{Mehrotra2023TreeOA}, membership inference operates at the distribution level and lacks explicit refusal boundaries. The leakage signal is statistical rather than deterministic, requiring the aggregation of logits over large batches to reveal discrepancies. This dependency on aggregated, implicit feedback makes instantaneous credit assignment for the agent's actions significantly harder than in scenarios with clear optimization targets; 

\textbf{(ii) Combinatorial Complexity of Strategy Space.} Existing handcrafted methods rely on expert-driven heuristics targeting specific statistical properties (e.g., entropy)~\cite{Carlini2021MembershipIA}. Automating this process requires the agent to navigate a vast combinatorial space of potential logits-level operations without prior knowledge of discriminative features. This immense search space, coupled with the heterogeneity of target model architectures, poses a significant challenge for efficient strategy discovery and adaptation.

\section{Method}

\subsection{Overview}

Figure~\ref{fig:automia} illustrates the overall architecture of AutoMIA, a framework designed to automate membership inference attacks via iterative self-exploration. Following the notation defined earlier, we use $t$ to index the iteration (round) and $i$ to index the $i$-th candidate strategy. The dynamic strategy library at iteration $t$ is denoted as $\mathcal{B}_t$, and the retrieved context from the previous round is a compact subset of strategies, $\mathcal{C}_t \subseteq \mathcal{B}_{t-1}$. The reflective guidance signal produced by the Guidance agent is denoted as $g_t$.

At each iteration, the AutoMIA agent proposes $K$ candidate strategies $\{(\textcolor{RubineRed}{s_t^{i}}, \textcolor{LightSlateGray}{\mathbf{p}_t^{i}})\}_{i=1}^{K}$, where $\textcolor{RubineRed}{s_t^{i}}$ denotes a high-level strategy specification (semantic description and mathematical formulation), and $\textcolor{LightSlateGray}{\mathbf{p}_t^{i}}$ is its associated logits-level runnable code. An example of the candidate strategy can be found in Fig.~\ref{fig:step_simi} (Right). Each candidate strategy is evaluated and summarized as a tuple $\textcolor{Tan}{r_t^{i}}$ (including three terms, detailed in Sec.~\ref{subsec:Strategy Library and Selection Mechanism}) and a composite score $Q(\textcolor{RubineRed}{s_t^{i}}, \textcolor{Tan}{r_t^{i}})$. The guidance step is written as $(g_t, \{\textcolor{CornflowerBlue}{\hat{s}_t^{i}}\}_{i=1}^{K}) \leftarrow \mathcal{H}(\cdot)$, where $\mathcal{H}(\cdot)$ denotes the Guidance agent, which outputs a natual language guidance $g_t$ and a categorized set of strategies $\{\textcolor{CornflowerBlue}{\hat{s}_t^{i}}\}_{i=1}^{K}$. Compared to the uncategorized/original version, the categorized version for each strategy additionally include a strong/weak label and some analysis. Concrete examples are provided in Appendix~\ref{app:Example for strategy library}. The strategy library then incorporates these categorized strategies for the next generation.

At the outset, the target model is queried on the target dataset containing both members and nonmembers to obtain the corresponding logits, which can be reused throughout the iterations without repeated computation. Starting from an empty repository, the strategy library gradually evolves into a knowledge base that supports subsequent strategy updates. In each iteration, the AutoMIA agent leverages $\mathcal{C}_t$ and $g_t$ from the strategy library and the Guidance agent respectively as its context to synthesize next round's candidate strategies and executable attack code, which is executed on the reusable logits within the \textit{Code Execution} module. The Guidance agent subsequently evaluates the outcomes and produces next round's reflective guidance. Finally, we log each newly generated strategy and its evaluation statistics to the strategy library, allowing the attack logic to improve via accumulated experience across iterations.

\subsection{Strategy Library and Selection Mechanism}
\label{subsec:Strategy Library and Selection Mechanism}

To facilitate stable and efficient traversal of the attack strategy space, we maintain a dynamic \textit{Strategy Library} $\mathcal{B}_t$, which archives generated strategies together with their empirical performance statistics (examples are provided in Appendix~\ref{app:Example for strategy library}). Each strategy is evaluated using a set of complementary metrics: Area Under the ROC Curve (AUC), Classification Accuracy (Acc), and True Positive Rate at a fixed False Positive Rate (TPR@5\%FPR) forming an evaluation tuple $r = (\mathrm{AUC}, \mathrm{Acc}, \mathrm{TPR})$.

To synthesize these distinct performance dimensions into a unified optimization objective, we aggregate them into a scalar \textbf{Composite Effectiveness Score}, denoted as $Q(s, r)$, via a weighted linear combination of the metrics tuple $r$ of a candidate strategy $s$. The scoring function $Q(s,r)$ can be formally defined as:
\begin{equation}
Q(s, r)
= w_{\mathrm{AUC}} \cdot \mathrm{AUC}
+ w_{\mathrm{Acc}} \cdot \mathrm{Acc}
+ w_{\mathrm{TPR}} \cdot \mathrm{TPR}.
\label{eq:score_func}
\end{equation}
where coefficients $w_{\mathrm{AUC}}$, $w_{\mathrm{Acc}}$, and $w_{\mathrm{TPR}}$ calibrate the relative importance of each metric (ablations are detailed in Sec.~\ref{subsec:ablation_Weights}). This scalarization prioritizes general discriminative power while strictly enforcing robustness in low false-positive regimes, thereby offering a faithful characterization of practical attack effectiveness.

During the exploration phase, we identify a recurrent challenge wherein the agent, driven by inherent stochasticity, may cyclically propose variations of strategies that yield consistently suboptimal results. This phenomenon, which we term \emph{inefficient exploration}, typically stems from unguided reasoning uncertainties and results in redundant computational expenditure without tangible performance convergence. To suppress inefficient exploration while alleviating the agent’s contextual memory burden, we adopt a fixed-size \emph{sliding window} mechanism for strategy selection. At each iteration $t$, instead of exposing the agent to the entire strategy library $\mathcal{B}_t$, only a compact subset of strategies $\mathcal{C}_t$ is provided as contextual input, as formally defined in Eq.~\ref{library_func}: 
\begin{equation}
\mathcal{C}_t =
\begin{cases}
\varnothing, & t = 0, \\[4pt]
\mathcal{B}_{t-1}, & t > 0 \ \text{and}\ |\mathcal{B}_{t-1}| \le w, \\[4pt]
\mathcal{C}_t^{+} \cup \mathcal{C}_t^{-}, & t > 0 \ \text{and}\ |\mathcal{B}_{t-1}| > w.
\end{cases}
\label{library_func}
\end{equation}
As the strategy library evolves over iterations, the composition of $\mathcal{C}_t$ varies accordingly with $t$, reflecting the progressively accumulated experience. This subset $\mathcal{C}_t$ consists of two categories of strategies, namely \emph{high-quality strategies}($\mathcal{C}_t^{+}$) with the highest composite scores $Q(s)$  and \emph{low-quality strategies}($\mathcal{C}_t^{-}$) with the lowest scores, their quantities determined by the size of the sliding window $w$ (The specific value can be found in Sec.~\ref{sec:implementation_details}). By jointly exposing representative successful and unsuccessful strategies, this design guides the agent toward promising strategy directions while helping it identify and avoid repeatedly sampling strategy patterns that have already demonstrated poor performance, thereby improving overall exploration efficiency by maintaining a focused and relevant reasoning context.

\begin{table*}[t!]
  \caption{AUC comparison of membership inference attacks under different text lengths ($L\in\{32,64\}$) on three vision--language models (LLaVA, MiniGPT-4, and LLaMAAdapter). Results are reported for representative baselines and our agent-generated strategy (Agent/Ours). We highlight the best, second-best, and third-best results in progressively lighter shades of blue, and mark the worst, second-worst, and third-worst results in progressively lighter shades of red.}

  \vspace{-3mm}
  \label{tab:three_models_text}
  \begin{center}
    \begin{small}
      \begin{sc}
        \resizebox{\textwidth}{!}{%
        \begin{tabular}{l l c c c c c c}
        \toprule
        \multirow{2}{*}{Metric} &
        & \multicolumn{2}{c}{LLaVA}
        & \multicolumn{2}{c}{MiniGPT-4}
        & \multicolumn{2}{c}{LLaMA Adapter} \\
        \cmidrule(lr){3-4}\cmidrule(lr){5-6}\cmidrule(lr){7-8}
        &
        & Text$_{\text{len}=32}$ & Text$_{\text{len}=64}$
        & Text$_{\text{len}=32}$ & Text$_{\text{len}=64}$
        & Text$_{\text{len}=32}$ & Text$_{\text{len}=64}$ \\
        \midrule

        \multicolumn{2}{l}{Perplexity}
        & 0.779 & \thirdbest{0.988} & \secondbest{0.702} & \thirdbest{0.823} & \secondbest{0.791} & 0.431 \\
        \multicolumn{2}{l}{Max Prob Gap}
        & \thirdworst{0.462} & \thirdworst{0.545} & 0.637 & \worst{0.418} & 0.583 & \secondbest{0.616} \\
        \midrule

        Min-$k$ Prob & Min-0\%
        & 0.522 & \secondworst{0.522} & 0.581 & 0.538 & 0.623 & \secondworst{0.366} \\
        & Min-10\%
        & \secondworst{0.461} & 0.883 & 0.585 & 0.668 & 0.658 & 0.375 \\
        & Min-20\%
        & 0.603 & 0.980 & 0.619 & 0.738 & 0.717 & 0.390 \\
        \midrule

        ModR\'enyi / Gap & $\alpha=0.5$
        & \secondbest{0.809} & 0.979 & 0.617 & 0.782 & 0.705 & 0.448 \\
        & $\alpha=1$
        & \thirdbest{0.808} & \secondbest{0.993} & \thirdbest{0.698} & 0.823 & \thirdbest{0.787} & 0.426 \\
        & $\alpha=2$
        & 0.779 & 0.963 & 0.540 & 0.712 & 0.656 & 0.441 \\
        \midrule

        R\'enyi ($\alpha=0.5$) & Max-0\%
        & 0.506 & \worst{0.514} & 0.524 & 0.651 & 0.654 & 0.382 \\
        & Max-10\%
        & \worst{0.458} & 0.776 & \worst{0.309} & 0.674 & 0.670 & 0.404 \\
        & Max-100\%
        & 0.564 & 0.835 & 0.611 & \secondbest{0.845} & 0.647 & \worst{0.365} \\
        \midrule

        R\'enyi ($\alpha=1$) & Max-0\%
        & 0.554 & 0.579 & 0.521 & 0.618 & 0.608 & 0.389 \\
        & Max-10\%
        & 0.566 & 0.809 & \secondworst{0.387} & 0.653 & 0.619 & 0.395 \\
        & Max-100\%
        & 0.554 & 0.750 & 0.617 & 0.802 & 0.674 & 0.419 \\
        \midrule

        R\'enyi ($\alpha=2$) & Max-0\%
        & 0.589 & 0.625 & 0.525 & \thirdworst{0.499} & 0.597 & 0.385 \\
        & Max-10\%
        & 0.606 & 0.787 & \thirdworst{0.488} & 0.605 & \thirdworst{0.581} & \thirdworst{0.369} \\
        & Max-100\%
        & 0.553 & 0.709 & 0.620 & 0.740 & 0.671 & 0.485 \\
        \midrule

        R\'enyi ($\alpha=\infty$) & Max-0\%
        & 0.601 & 0.638 & 0.522 & \secondworst{0.474} & \worst{0.575} & 0.411 \\
        & Max-10\%
        & 0.618 & 0.763 & 0.497 & 0.592 & \secondworst{0.578} & 0.378 \\
        & Max-100\%
        & 0.557 & 0.694 & 0.621 & 0.701 & 0.672 & \thirdbest{0.522} \\
        \midrule

        AUTOMIA (Ours)
        & DeepSeek-V3.2-Reasoner
        & \best{0.810} & \best{0.994}
        & \best{0.824} & \best{0.891}
        & \best{0.828} & \best{0.778} \\
        \bottomrule
        \end{tabular}
        }
      \end{sc}
    \end{small}
  \end{center}
 \vspace{-4mm}
\end{table*}

Fig.~\ref{fig:automia} illustrates how the retrieved strategy subset $\mathcal{C}_t$ and the Guidance agent’s evaluation of the prior strategy jointly form the feedback signal that drives the AutoMIA agent’s next-round generation. Collectively, the exemplar strategies and the diagnostic feedback constitute a dense, informative conditioning context that steers the agent's reasoning during the subsequent generation cycle. Consequently, the strategy library evolves beyond a passive storage role, serving as an active control component that dynamically balances exploration and exploitation under noisy conditions. Furthermore, by coupling weighted multi-metric evaluation with a token-efficient sliding window, this design minimizes redundant trials and stabilizes the agent’s iterative refinement trajectory under strict computational constraints.

\subsection{AutoMIA and Guidance agents}
\label{subsec:agent_reasoning}

The AutoMIA agent coordinates the generation, execution, and iterative refinement of attack strategies through an explicit reasoning and decision-making process.
In contrast to conventional approaches that optimize a predefined objective, the agent proceeds iteratively under feedback, with each action conditioned on the growing execution trace and corresponding evaluation signals. We now describe the key components of AutoMIA, including strategy synthesis, execution and evaluation, and guidance-driven library updates.

\noindent\textbf{Strategy synthesis.}
The AutoMIA agent performs high-level reasoning to determine its next action by proposing a set of candidate MIA strategies. Conditioned on the retrieved context $\mathcal{C}_t \subseteq \mathcal{B}_{t-1}$ and the previous-round guidance $g_{t-1}$ from the Guidance agent, the agent synthesizes $K$ candidate strategies $\{(\textcolor{RubineRed}{s_t^{i}}, \textcolor{LightSlateGray}{\mathbf{p}_t^{i}})\}_{i=1}^{K}$, where each $\textcolor{RubineRed}{s_t^{i}}$ specifies an abstract attack strategy and $\textcolor{LightSlateGray}{\mathbf{p}_t^{i}}$ is its executable logits-level instantiation on the target model.

\noindent\textbf{Execution and evaluation.}
The agent’s decision-making policy is not governed by formal reward maximization; rather, it is iteratively steered by empirical feedback obtained through execution and evaluation. Specifically, as we've mentioned earlier, the target dataset $\mathcal{D}$ is firstly queried on the target model $M$ to collect the reusable logits $\mathbf{o}$. For each candidate strategy, its executable attack code $\textcolor{LightSlateGray}{\mathbf{p}_t^{i}}$ is applied to $\mathbf{o}$ to produce per-sample membership scores. These scores are then used to compute standard evaluation metrics (AUC, Accuracy, and $\mathrm{TPR@5\%FPR}$), with decisions made via Eq.~\ref{eq:membership_score}. We summarize the value of these three metrics as an evaluation tuple $\textcolor{Tan}{r_t^{i}}$ for strategy $i$ in the $t$-th iteration. Finally, following Eq.~\ref{eq:score_func}, we aggregate the metrics tuple $\textcolor{Tan}{r_t^{i}}$ into a scalar \textbf{Composite Effectiveness Score} $Q(\textcolor{RubineRed}{s_t^{i}}, \textcolor{Tan}{r_t^{i}})$ via a weighted linear combination, and use this scalar feedback to guide subsequent strategy refinement.

\noindent\textbf{Guidance and library update.}
After execution, the collection of evaluation signals $\{\textcolor{Tan}{r_t^{i}}, Q(\textcolor{RubineRed}{s_t^{i}}, \textcolor{Tan}{r_t^{i}})\}_{i=1}^{K}$ is forwarded to the Guidance agent to get its guidance for the next iteration $g_t$ and the categorized strategies in the current iteration $\{\textcolor{CornflowerBlue}{\hat{s}_t^{i}}\}_{i=1}^{K}$. This step can be formally defined as:
\begin{equation}
(g_t, \{\textcolor{CornflowerBlue}{\hat{s}_t^{i}}\}_{i=1}^{K}) \leftarrow \mathcal{H}\!\left(\{\textcolor{Tan}{r_t^{i}},\textcolor{RubineRed}{s_t^{i}}, Q(\textcolor{RubineRed}{s_t^{i}}, \textcolor{Tan}{r_t^{i}})\}_{i=1}^{K}\right) .   
\label{eq:guidance_feedback}
\end{equation}
The strategy library is then updated by incorporating the categorized strategies together with their evaluation statistics and reflective guidance signals:
\begin{equation}
\mathcal{B}_t = \mathcal{B}_{t-1}  \cup \mathcal{U}\!\left(\{\textcolor{CornflowerBlue}{\hat{s}_t^{i}}, \textcolor{Tan}{r_t^{i}}, Q(\textcolor{RubineRed}{s_t^{i}}, \textcolor{Tan}{r_t^{i}})\}_{i=1}^{K}\right),
\label{eq:lib_update}
\end{equation}
where $\mathcal{U}(\cdot)$ denotes the procedure for formatting useful information into the strategy library. Overall, the AutoMIA agent and the Guidance agent together form a closed-loop decision-making entity that follows a perception--reasoning--action--reflection cycle, enabling systematic and effective exploration of the broad and noisy attack space.

\section{Experiment}
\subsection{Experimental Setup}
\label{sec:Experimental Setup}

\begin{table*}[t]
\caption{VL-MIA AUC Comparison on DALL$\cdot$E and Fliker with LLaVA as the victim model. `img' indicates the logits slice corresponding to image embedding, `inst' indicates the instruction slice, `desp' the generated description slice, and `inst+desp' is the concatenation of the instruction slice and description slice. For the image slice, target-based MIAs are not applicable due to the absence of ground-truth token IDs, and the corresponding results are therefore reported as N/A. We highlight the best, second-best, and third-best results in progressively lighter shades of blue, and mark the worst, second-worst, and third-worst results in progressively lighter shades of red.}

  \vspace{-3mm}
  \label{tab:vlmia_merged_llava_compact}
  \begin{center}
    \begin{small}
      \begin{sc}
        \resizebox{\textwidth}{!}{%
        \begin{tabular}{l l c c c c c c c c}
        \toprule
        \multirow{2}{*}{Metric} &
        & \multicolumn{4}{c}{DALL$\cdot$E (LLaVA)}
        & \multicolumn{4}{c}{Fliker (LLaVA)} \\
        \cmidrule(lr){3-6}\cmidrule(lr){7-10}
        &
        & img & inst & desp & inst+desp
        & img & inst & desp & inst+desp \\
        \midrule

        \multicolumn{2}{l}{Perplexity}
        & N/A & \worst{0.337} & \thirdbest{0.567} & 0.448
        & N/A & 0.378 & 0.662 & 0.554 \\
        \multicolumn{2}{l}{Max Prob Gap}
        & 0.529 & 0.578 & \best{0.598} & \thirdbest{0.603}
        & 0.579 & 0.603 & \thirdworst{0.645} & 0.646 \\
        \multicolumn{2}{l}{Aug-KL}
        & \secondworst{0.432} & 0.462 & \worst{0.523} & 0.504
        & 0.605 & 0.538 & \worst{0.476} & 0.496 \\
        \midrule

        Min-$k$ Prob & Min-0\%
        & N/A & 0.481 & 0.556 & 0.481
        & N/A & \worst{0.358} & 0.647 & \worst{0.358} \\
        & Min-10\%
        & N/A & 0.481 & 0.561 & \secondworst{0.424}
        & N/A & \worst{0.358} & 0.667 & \thirdworst{0.390} \\
        & Min-20\%
        & N/A & 0.434 & 0.560 & \worst{0.352}
        & N/A & 0.374 & 0.668 & \secondworst{0.370} \\
        \midrule

        ModR\'enyi / Gap & $\alpha=0.5$
        & N/A & \thirdworst{0.359} & 0.563 & 0.525
        & N/A & \thirdworst{0.368} & 0.646 & 0.609 \\
        & $\alpha=1$
        & N/A & \secondworst{0.341} & 0.563 & \thirdworst{0.425}
        & N/A & \secondworst{0.359} & 0.654 & 0.499 \\
        & $\alpha=2$
        & N/A & 0.383 & 0.564 & 0.539
        & N/A & 0.370 & \secondworst{0.640} & 0.605 \\
        \midrule

        R\'enyi ($\alpha=0.5$) & Max-0\%
        & 0.553 & 0.598 & 0.557 & 0.598
        & \secondworst{0.513} & 0.689 & 0.682 & 0.689 \\
        & Max-10\%
        & 0.622 & 0.598 & 0.559 & \secondbest{0.644}
        & \thirdworst{0.554} & 0.689 & 0.687 & 0.718 \\
        & Max-100\%
        & \worst{0.421} & 0.605 & \secondbest{0.575} & 0.582
        & \best{0.701} & \secondbest{0.726} & \secondbest{0.707} & 0.722 \\
        \midrule

        R\'enyi ($\alpha=1$) & Max-0\%
        & 0.549 & 0.569 & \thirdworst{0.549} & 0.575
        & \worst{0.496} & 0.707 & 0.680 & \thirdbest{0.724} \\
        & Max-10\%
        & 0.666 & 0.569 & 0.557 & 0.586
        & 0.619 & 0.707 & 0.694 & \best{0.739} \\
        & Max-100\%
        & \thirdworst{0.470} & \secondbest{0.638} & 0.566 & 0.586
        & \best{0.701} & \thirdbest{0.720} & \thirdbest{0.696} & 0.716 \\
        \midrule

        R\'enyi ($\alpha=2$) & Max-0\%
        & 0.593 & 0.549 & \secondworst{0.543} & 0.558
        & 0.582 & 0.682 & 0.666 & 0.700 \\
        & Max-10\%
        & \secondbest{0.705} & 0.549 & 0.551 & 0.575
        & 0.617 & 0.682 & 0.681 & 0.719 \\
        & Max-100\%
        & 0.526 & \thirdbest{0.606} & 0.564 & 0.579
        & \thirdbest{0.680} & 0.694 & 0.676 & 0.697 \\
        \midrule

        R\'enyi ($\alpha=\infty$) & Max-0\%
        & 0.625 & 0.560 & 0.556 & 0.568
        & 0.586 & 0.647 & 0.647 & 0.671 \\
        & Max-10\%
        & \thirdbest{0.698} & 0.560 & 0.561 & 0.582
        & 0.593 & 0.647 & 0.667 & 0.696 \\
        & Max-100\%
        & 0.545 & 0.588 & \thirdbest{0.567} & 0.580
        & 0.668 & 0.673 & 0.662 & 0.683 \\
        \midrule

        AUTOMIA (Ours) & DeepSeek-V3.2-Reasoner
        & \best{0.787} & \best{0.663} & \best{0.598} & \best{0.653}
        & \secondbest{0.700} & \best{0.729} & \best{0.715} & \secondbest{0.734} \\
        \bottomrule
        \end{tabular}
        }
      \end{sc}
    \end{small}
  \end{center}
\vspace{-6mm}
\end{table*}

\noindent\textbf{Datasets.}
We evaluate AutoMIA on three benchmark datasets~\cite{Li2024MembershipIA} for membership inference attacks against large vision-language models (denoted as VL-MIA, short for \emph{Vision--Language Model Membership Inference Attack}): VL-MIA/Text, VL-MIA/DALL$\cdot$E, and VL-MIA/Flickr. 
VL-MIA/Text targets the instruction-tuning stage, where member texts are sampled from instruction-tuning data with descriptive answers of fixed lengths, while non-member texts are generated by GPT-4 using matched questions, images, and text lengths.
VL-MIA/DALL$\cdot$E focuses on the image modality, constructing paired member and non-member samples by sampling training images shared across multiple VLLMs and generating corresponding non-member images via DALL$\cdot$E using BLIP captions.
VL-MIA/Flickr uses MS COCO images as member data and Flickr images uploaded after Jan.~1,~2024 as non-members, and additionally includes corrupted versions of member images to simulate realistic deployment conditions.

\noindent\textbf{Baselines.} 
We compare our framework against a comprehensive suite of state-of-the-art handcrafted metrics commonly used in membership inference. 
We strictly follow the setup in prior work~\cite{Li2024MembershipIA} and include: 
(i) \textbf{Perplexity}~\cite{yeom2018privacy}, which measures the model's prediction uncertainty on the target sample; 
(ii) \textbf{Max Probability Gap}, which calculates the difference between the highest and second-highest token probabilities; 
and (iii) \textbf{Min-$k$\% Prob}~\cite{shi2023detecting}, a state-of-the-art method for LLMs that focuses on the average likelihood of the $k$\% tokens with the lowest probability. 
Furthermore, we incorporate the recently proposed \textbf{Rényi} and \textbf{ModRényi} families of metrics~\cite{Li2024MembershipIA}, which generalize entropy-based attacks using Rényi divergence. 
For these, we evaluate multiple configurations with varying orders ($\alpha \in \{0.5, 1, 2, \infty\}$) and pooling strategies (e.g., Max-$k$\%) to ensure a robust comparison against the strongest existing heuristics.

\begin{figure*}[t]
    \centering
    \includegraphics[width=\textwidth]{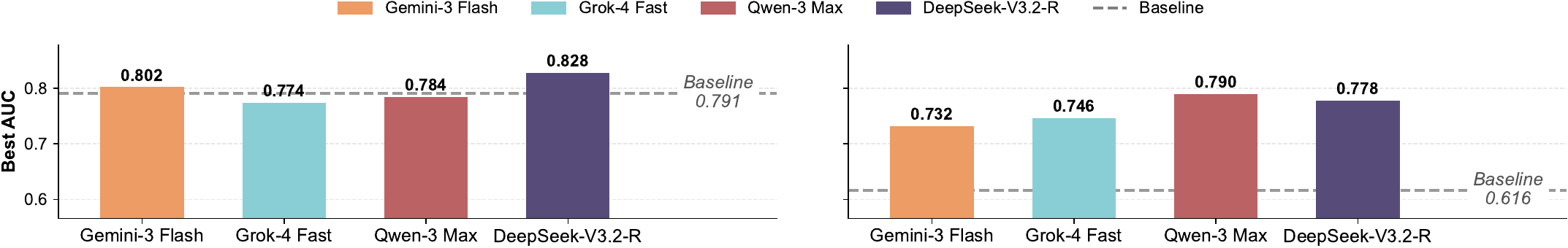}
    \caption{
        Ablation on Agent Backbone. Performance comparison of AutoMIA driven by different VLM backbones (Gemini 3 Flash, Grok 4.1 Fast, Qwen3-Max, and DeepSeek-V3.2-Reasoner) on LLaMA-Adapter.
    }
    \label{fig:ablation_coreLLM}
\end{figure*}
\begin{figure*}[t]
    \centering
    \includegraphics[width=0.9\textwidth]{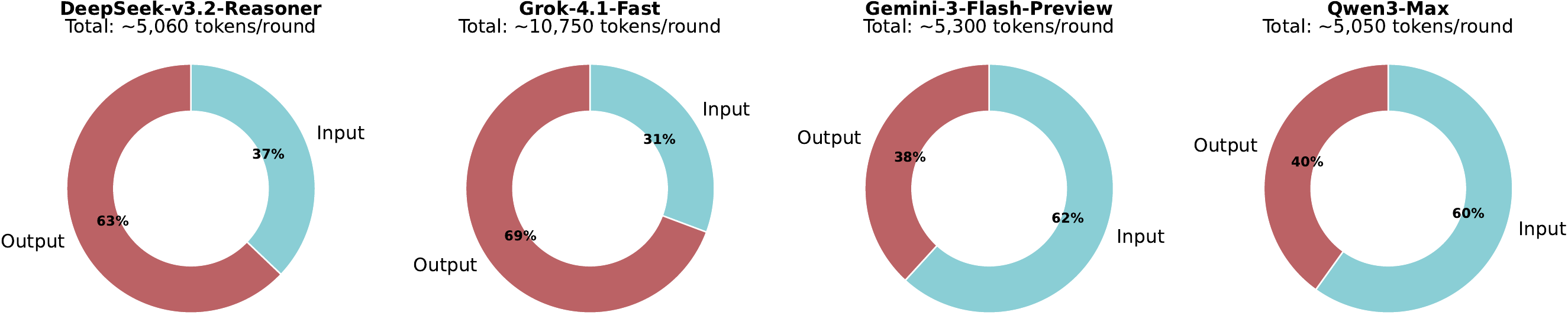}
    \caption{
            Token Consumption Figure: Input vs Output for Different VLM Models. Total tokens per round are indicated for each model. Red represents the output tokens, and blue represents the input tokens.
    }
    \label{fig:tokens_consumption}
    \vspace{-4mm}
\end{figure*}
\noindent\textbf{Target Models.}
To ensure rigorous comparability with prior baselines, we align our target model selection with the well-established protocols~\cite{Li2024MembershipIA}. Specifically, we evaluate three representative open-source Large Vision-Language Models (LVLMs): MiniGPT-4~\cite{zhu2023minigpt}, LLaVA-1.5~\cite{liu2024improved}, and LLaMA-Adapter~\cite{zhang2023llama}. These models were selected for their architectural diversity, the availability of transparent training pipelines, and their established role as standard baselines in membership inference literature. All three models adhere to a multi-stage training paradigm, encompassing unimodal pre-training, multimodal alignment, and instruction tuning. Consistent with the dataset configuration, we adopt the member/non-member split in~\cite{li2024membership}, strictly utilizing instruction-tuning responses as member data and GPT-4 synthesized counterparts under identical image-instruction pairs as non-member data. This standardized setup effectively isolates the experimental variables, allowing us to attribute performance gains directly to the automated strategy evolution of AutoMIA rather than discrepancies in target model configurations.

\noindent\textbf{Attack Settings and Access Assumptions.}
All experiments are conducted under a grey-box threat model.
The agent has no access to model parameters or training data, but can observe logits or confidence-related outputs returned by the target model.
This setting reflects realistic deployment scenarios for large vision--language models and is consistent with prior work on grey-box MIA evaluation.

\noindent\textbf{Implementation and Strategy Details.}
\label{sec:implementation_details}
All experiments are implemented in PyTorch and conducted on a single NVIDIA RTX~4090 GPU with 24GB memory.
The temperature of all models is fixed to 0.6, and each experimental configuration is executed for ten rounds.
Experiments are conducted consistently across VL-MIA/Text, VL-MIA/DALL$\cdot$E, and VL-MIA/Flickr under the same experimental protocol.
The strategy library is initialized as empty at the beginning of the experiments.
In the first round, the agent freely explores candidate attack metrics without prior constraints.
After each round, strategies are evaluated using a weighted composite score
\(
S = 0.6\,\mathrm{AUC}
+ 0.3\,\mathrm{Acc}
+ 0.1\,\mathrm{TPR@5\%FPR}.
\)
Based on the score distribution, strategies are dynamically categorized into strong, mid, and weak groups using the 70th and 30th percentiles.
The best-performing and worst-performing strategies are stored in the strategy library.
In subsequent rounds, three strong and two weak strategies are selected to guide further exploration, using a sliding window of size $w = 5$ to analyze the most recent strategies.

\subsection{Overall Performance Comparison}

We compare AutoMIA with a wide range of representative membership inference metrics across three vision--language models and multiple evaluation settings.
Tables~\ref{tab:three_models_text} to \ref{tab:vlmia_merged_llava_compact} report AUC scores on text-based, image-based, and multimodal benchmarks, respectively.

\textbf{Text-based MIA.}
As shown in Table~\ref{tab:three_models_text}, existing handcrafted metrics exhibit highly inconsistent performance across models and text lengths.
While certain metrics achieve strong results under specific configurations (e.g., long text or particular architectures), their effectiveness degrades substantially when the setting changes.
In contrast, AutoMIA consistently achieves near-optimal performance across all models and text lengths, outperforming the strongest baseline by a clear margin.
This result indicates that automated strategy discovery is substantially more robust than relying on fixed, manually designed metrics.

\textbf{Image and multimodal MIA.}
Tables~\ref{tab:vlmia_merged_llava_compact} further evaluate performance on image-centric and multimodal benchmarks.
Across both Flickr-based and DALL$\cdot$E-generated datasets, handcrafted metrics show large variance depending on which input components are used (image, instruction, description, or their combinations).
No single baseline metric generalizes well across models or modalities.
In contrast, AutoMIA consistently ranks among the top-performing methods and frequently achieves the best AUC across different modality compositions, demonstrating strong adaptability to heterogeneous output structures.

\begin{figure*}[ht]
  \centering

  \begin{minipage}[t]{0.66\textwidth}\vspace{0pt}
    \centering
    \includegraphics[width=\linewidth]{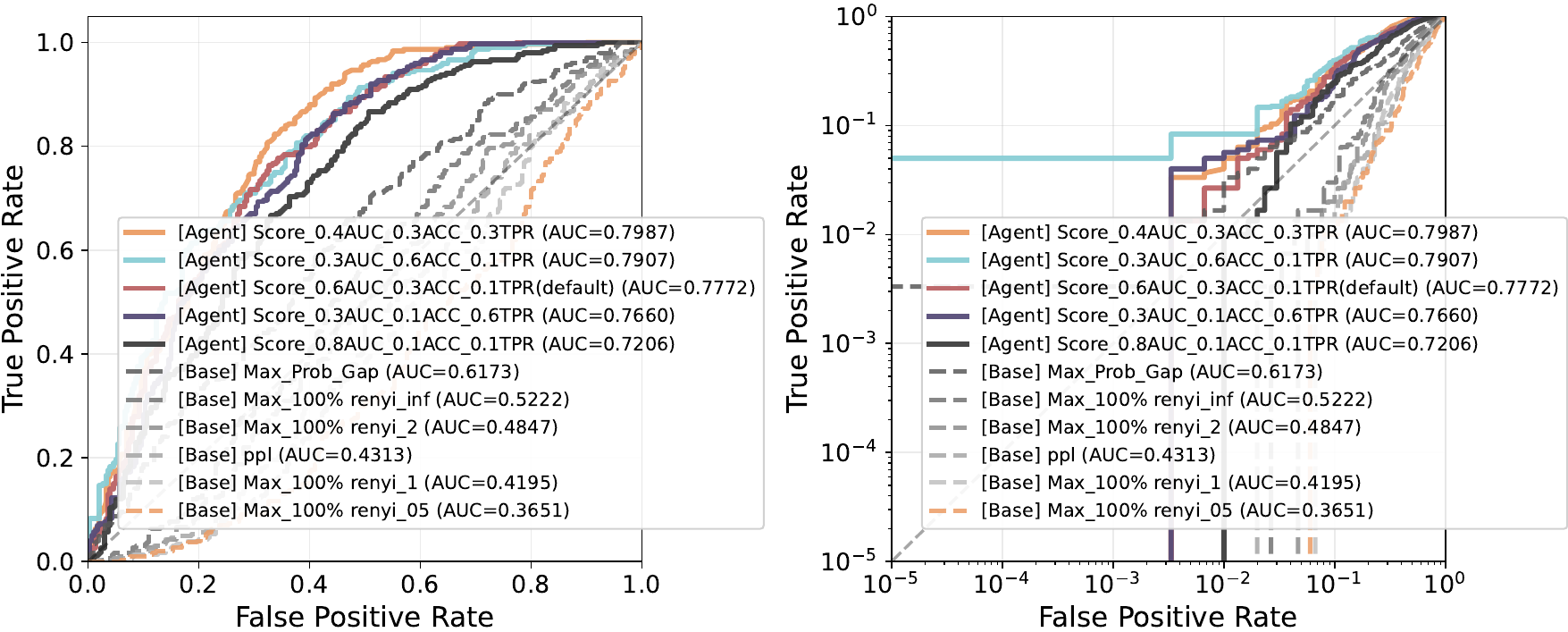}
    \captionof{figure}{Ablation study on the impact of scoring function weights for AutoMIA. The left panel compares ROC curves with linear FPR for different scoring configurations, including agent-generated strategies and baselines. The right panel shows the same comparison with logarithmic FPR, highlighting the sensitivity-specificity trade-off.}
    \label{fig:auc_tpr}
  \end{minipage}
  \hfill
  \begin{minipage}[t]{0.32\textwidth}\vspace{0pt}
    \centering
    \includegraphics[width=\linewidth]{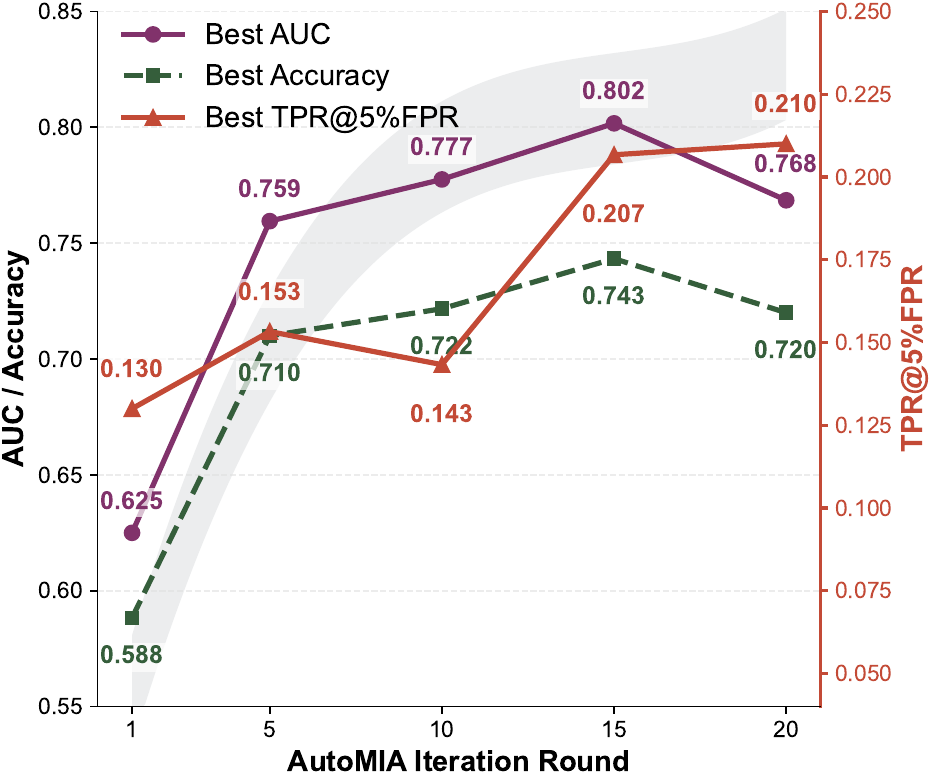}
    \captionof{figure}{Performance comparison of AutoMIA under different iteration rounds. The figure shows the best AUC, accuracy, and TPR@5\%FPR achieved across 20 iterations.}
    \label{fig:ablation_iterations}
  \end{minipage}

  \vspace{-4mm} 
\end{figure*}
Taken together, these results reveal a clear pattern:
while existing MIA methods are highly sensitive to model architecture, modality, and evaluation setting, AutoMIA maintains stable and competitive performance across all tested scenarios.
This robustness stems from its ability to automatically explore, evaluate, and refine attack strategies, rather than committing to a fixed metric design.
The overall comparison highlights the advantage of agent-driven membership inference in addressing the growing diversity of modern vision--language models.

\section{Ablation Study}
\label{sec:ablation}

\subsection{Impact of Agent Backbone}
\label{subsec:ablation_backbone}

To assess the dependency of AutoMIA on specific reasoning capabilities, we evaluate the framework using four distinct LLM backbones: Gemini 3 Flash~\cite{team2024gemini}, Grok 4.1 Fast~\cite{xai2025grok41}, Qwen3-Max~\cite{bai2023qwen}, and our default DeepSeek-V3.2-Reasoner. 
As shown in Figure~\ref{fig:ablation_coreLLM}, while the choice of backbone introduces minor variations in peak performance, AutoMIA consistently synthesizes high-efficacy strategies across all evaluated generators.
Specifically, under the shorter text setting ($L=32$), all agents converge to a comparable high-AUC regime, suggesting that the iterative self-exploration mechanism effectively compensates for differences in base reasoning capabilities.
Although increasing the input length to $L=64$ introduces moderate performance fluctuations due to the harder extraction task, the framework maintains strong effectiveness regardless of the proprietary model used, confirming that attack success is primarily driven by the closed-loop optimization process rather than the specific parametric knowledge of the backbone.
\\
In addition to effectiveness, we analyze the per-round token consumption of different backbones to assess the practical cost of running AutoMIA (Figure~\ref{fig:tokens_consumption}). 
Among the four generators, Gemini 3 Flash and Qwen3-Max show the most favorable token consumption patterns: their total tokens per round are comparable to DeepSeek-V3.2-Reasoner and substantially lower than Grok 4.1 Fast, while allocating a smaller fraction of tokens to model outputs. 
Since output tokens are typically billed at a higher rate than input tokens, this reduced output share leads to lower overall cost. 
Overall, Gemini 3 Flash and Qwen3-Max emerge as attractive backbones for large-scale exploration, balancing strong strategy quality with lower generation overhead.

\subsection{Impact of Exploration Rounds}
\label{subsec:ablation_rounds}

We further investigate the temporal dynamics of strategy evolution by tracking attack performance over increasing exploration rounds on the LLaMA-Adapter target (Text$_{\text{len}=64}$).
As illustrated in Figure~\ref{fig:ablation_iterations}, the optimization process exhibits a clear convergence trajectory. 
In the initial iterations (rounds 1--5), the agent achieves substantial performance gains, indicating that the closed-loop feedback effectively steers exploration toward promising regions of the attack space.
Performance continues to improve and typically peaks around the 15th round, where the accumulated strategy library and guidance signals enable the refinement and consolidation of effective attack patterns.
Beyond this point, extending the computational budget yields diminishing marginal returns as the performance metrics stabilize. 
This trajectory demonstrates that AutoMIA is sample-efficient, capable of reaching near-optimal performance within a reasonable budget (approx. 15 rounds) while maintaining stability over extended exploration.

\subsection{Impact of Scoring Function Weights}
\label{subsec:ablation_Weights}

We conduct an ablation study on the scoring function $Q(s,r)$ for the \textsc{LLaMA Adapter} (text length 64) to examine how different weighting configurations influence the strategies synthesized by the agent. 
Across all variants, the strategies generated by the AutoMIA agent consistently outperform handcrafted baselines, highlighting the effectiveness of jointly leveraging multiple evaluation signals. 
We find that shifting the emphasis toward a single criterion leads to strategies that favor either localized sensitivity in restricted operating regions or smoother but less discriminative global behavior. 
In contrast, the default configuration achieves a more balanced trade-off, maintaining stable separation across the ROC curve while preserving sensitivity under low false positive rate (FPR) constraints. 
These trends are consistently observed across both linear and logarithmic FPR visualizations, as shown in Figure~\ref{fig:auc_tpr}.

\begin{table*}[t]
\caption{Generalizability of \textbf{top AutoMIA strategies} under a 50\% validation / 50\% hold-out test split.}
\vspace{-3mm}
\label{tab:heldout_generalization}
\begin{center}
\begin{small}
\begin{sc}
\resizebox{0.8\textwidth}{!}{%
\begin{tabular}{lccc|ccc}
\toprule
\multirow{2}{*}{Generated Strategy} 
& \multicolumn{3}{c}{Validation Set (50\%)} 
& \multicolumn{3}{c}{Hold-out Set (50\%)} \\
\cmidrule(lr){2-4} \cmidrule(lr){5-7}
& AUC & Acc & TPR@5\% & AUC & Acc & TPR@5\% \\
\midrule

True-Token Probability Momentum
& 0.784 & 0.723 & 0.152
& \best{0.741} & 0.699 & 0.096 \\

True-Token Probability Consistency
& 0.784 & 0.724 & 0.164
& 0.738 & \best{0.706} & 0.104 \\

Probability Curvature Sign Consistency
& \best{0.792} & \best{0.751} & 0.133
& 0.735 & 0.694 & 0.082 \\

True-Token Relative-Confidence Momentum
& \best{0.792} & 0.733 & \best{0.182}
& 0.735 & 0.703 & \best{0.170} \\

True-Token Neighborhood Cohesion
& 0.773 & 0.727 & 0.176
& 0.663 & 0.628 & 0.096 \\

\bottomrule
\end{tabular}
}
\end{sc}
\end{small}
\end{center}
\vspace{-6mm}
\end{table*}

\begin{table}[t]
\caption{Performance comparison on the OLMo near-IID evaluation setting. We report the best AutoMIA strategy and the average performance of the top-5 AutoMIA strategies, together with representative baseline methods.}
\vspace{-3mm}
\label{tab:olmo_near_iid_results}
\begin{center}
\begin{small}
\begin{sc}
\resizebox{0.8\columnwidth}{!}{%
\begin{tabular}{lccc}
\toprule
\textbf{Method} & \textbf{AUC} & \textbf{Acc} & \textbf{TPR@5\%} \\
\midrule

\textbf{Best AutoMIA}
& \best{0.723} & \best{0.688} & \best{0.240} \\

\textbf{Top-5 AutoMIA}
& \secondbest{0.716} & \secondbest{0.678} & \thirdbest{0.207} \\

\midrule

Max\_100\% renyi\_05
& \secondbest{0.716} & \thirdbest{0.674} & \secondbest{0.216} \\

Max\_100\% renyi\_1
& 0.676 & 0.648 & 0.138 \\

Max\_100\% renyi\_2
& 0.642 & 0.618 & 0.130 \\

Max\_100\% renyi\_inf
& 0.633 & 0.612 & 0.138 \\

ppl
& \thirdbest{0.687} & 0.654 & 0.190 \\

Modified\_entropy
& 0.689 & 0.653 & 0.174 \\

Modified\_renyi\_05
& 0.643 & 0.627 & 0.114 \\

Modified\_renyi\_2
& 0.609 & 0.602 & \worst{0.098} \\

Max\_0\% renyi\_05
& \thirdworst{0.573} & \thirdworst{0.575} & \secondworst{0.100} \\

Max\_0\% renyi\_1
& \secondworst{0.562} & \secondworst{0.562} & \thirdworst{0.106} \\

Max\_0\% renyi\_2
& \worst{0.561} & \worst{0.561} & 0.110 \\

\bottomrule
\end{tabular}
}
\end{sc}
\end{small}
\end{center}
\vspace{-6mm}
\end{table}

\subsection{Evaluation under a Near-IID Setting.}
A common challenge in membership inference attacks (MIA) is that distribution shift between member and non-member data may lead to overestimated performance~\cite{das2025blind,meeus2025sok}. To mitigate this issue, we reconstruct the evaluation under a stricter near-IID setting.

Specifically, we adopt the open-source model OLMo-3-Instruct-7B-SFT~\cite{olmo2025olmo} and build the dataset from Dolma 3. Member samples are drawn from \texttt{dolma3\_mix-6T}, while non-member samples are drawn from the same source (\texttt{dolma3\_pool}) but excluded from training. We randomly sample 500 members and 500 non-members, control the text length to 64, and apply identical preprocessing. This keeps the two sets aligned in source and format, differing mainly in membership, and thus reduces cross-distribution artifacts such as synthetic bias or temporal shift. We further use random sampling and manual inspection to verify that no obvious structural differences (e.g., temporal or stylistic patterns) are present, suggesting that the constructed dataset approximately satisfies the IID assumption.

Under this stricter setting, the agent-discovered strategies still consistently outperform prior baselines, suggesting that the improvement comes from genuine memorization signals rather than dataset artifacts. In particular, the best discovered strategy surpasses the strongest baseline across all metrics, especially under low-FPR evaluation (TPR@5\%FPR: 0.240 vs. 0.216). Although the overall performance is moderately lower due to the increased difficulty of the near-IID setting, the method retains a clear advantage, indicating that the discovered attack signals are robust and transferable rather than dataset-specific.

\subsection{Unseen Data Generalizability (Held-out Test Split).}
To examine whether the proposed framework captures transferable privacy leakage patterns rather than overfitting to specific member/non-member instances, we further evaluate it under a held-out test protocol. Specifically, the dataset is divided into a 50\% validation split, used exclusively for strategy search and refinement, and a 50\% hold-out test split, used only for final evaluation on unseen data.

We observe that the top strategies discovered on the validation split generalize well to the hold-out test split, with only a moderate performance drop on unseen data. Despite this degradation, the hold-out AUCs remain substantially above random guessing and competitive with strong static baselines. These findings suggest that AutoMIA captures transferable statistical characteristics of model memorization rather than overfitting to dataset-specific artifacts.

\subsection{Impact of Guidance agent on Metric Exploration}
\label{subsec:guidance_ablation}

We study the role of the Guidance Agent in AutoMIA through an ablation experiment that removes it from the closed-loop discovery pipeline. In this setting, the agent still generates executable logits-level strategies based on prior results, but no longer receives explicit reflections or exploration suggestions.

As shown in Table~\ref{tab:guidance_ablation}, removing the Guidance Agent leads to a consistent performance drop across different text lengths. This trend indicates that the effectiveness of AutoMIA depends not only on executable strategy generation, but also on feedback-driven exploration. We attribute this difference to the difficulty of searching over a large and highly compositional metric space. Without guidance, the agent must explore candidate logit transformations with little directional bias, which makes the search process less efficient and less stable. By contrast, the Guidance Agent leverages evaluation feedback to suggest more promising directions, thereby improving the quality of exploration and accelerating convergence toward effective metrics.

\begin{table}[t]
\caption{Ablation study on the effect of the guidance agent in AutoMIA under different text lengths.}
\vspace{-3mm}
\label{tab:guidance_ablation}
\begin{center}
\begin{small}
\begin{sc}
\resizebox{\columnwidth}{!}{%
\begin{tabular}{ccccc}
\toprule
\textbf{Text Length} & \textbf{Method} & \textbf{AUC} & \textbf{Acc} & \textbf{TPR@5\%} \\
\midrule

32 
& w/o Guidance 
& \secondbest{0.709} & \secondbest{0.660} & \secondbest{0.147} \\

32 
& AutoMIA 
& \best{0.828} & \best{0.782} & \best{0.177} \\

\midrule

64 
& w/o Guidance 
& \secondbest{0.654} & \secondbest{0.623} & \secondbest{0.073} \\

64 
& AutoMIA 
& \best{0.787} & \best{0.722} & \best{0.143} \\

\bottomrule
\end{tabular}
}
\end{sc}
\end{small}
\end{center}
\vspace{-6mm}
\end{table}

\section{Conclusion}

In this work, we proposed AutoMIA, an agent-driven framework that reframes grey-box membership inference against vision--language models as an automated strategy generation and execution process. By enabling an agent to iteratively explore, evaluate, and refine logits-level attack strategies through closed-loop feedback, AutoMIA reduces reliance on handcrafted heuristics while remaining model-agnostic. Experiments across multiple vision--language models and datasets demonstrate that AutoMIA can adaptively explore and generate attack strategies tailored to each specific setting, achieving strong performance across diverse experimental conditions. More broadly, our work highlights the potential of agentic approaches for scalable and systematic privacy evaluation in large foundation models.

\bibliography{example_paper}
\bibliographystyle{icml2026}

\newpage
\appendix
\onecolumn

\section{Additional Experimental Results}
\label{app:additional_results}

In the main body of this paper, we primarily utilized the Area Under the ROC Curve (AUC) to benchmark membership inference performance, as it provides a threshold-independent measure of discriminative power. However, to offer a more holistic evaluation of privacy risks under different operating conditions, we present supplementary performance metrics in this appendix. Specifically, we report:

\begin{itemize}
    \item \textbf{Classification Accuracy (Acc):} Reflects the overall correctness of the attack when using an optimal threshold (maximized Youden's J statistic). This metric indicates the average success rate of the adversary in distinguishing members from non-members.
    \item \textbf{True Positive Rate at 5\% False Positive Rate (TPR@5\%FPR):} Measures the attack's sensitivity in a high-precision regime. This metric is critical for evaluating scenarios where the adversary requires high confidence and tolerates very few false alarms.
\end{itemize}

The following subsections detail these metrics for both text-based and multimodal benchmarks.

\subsection{Results on Text-Based Benchmarks}

Tables~\ref{tab:Acc_three_models_text} and \ref{tab:three_models_text_tpr5} present the Accuracy and TPR@5\%FPR comparisons, respectively, for the \textsc{VL-MIA/Text} dataset across LLaVA, MiniGPT-4, and LLaMA-Adapter. The results reinforce our findings from the main text: while handcrafted baselines like Perplexity and Min-$k$\% Prob exhibit significant volatility across different models and text lengths, \textsc{AutoMIA} consistently maintains high performance metrics, demonstrating superior robustness.

\begin{table}[h]
  \caption{Accuracy comparison of membership inference attacks under different text lengths ($L \in \{32, 64\}$) on three vision--language models (LLaVA, MiniGPT-4, and LLaMAAdapter). We highlight the best, second-best, and third-best results in progressively lighter shades of blue, and mark the worst, second-worst, and third-worst results in progressively lighter shades of red.}
  \label{tab:Acc_three_models_text}
  \begin{center}
    \begin{small}
      \begin{sc}
        \resizebox{\textwidth}{!}{%
        \begin{tabular}{l l c c c c c c}
        \toprule
        \multirow{2}{*}{Metric} &
        & \multicolumn{2}{c}{LLaVA}
        & \multicolumn{2}{c}{MiniGPT-4}
        & \multicolumn{2}{c}{LLaMAAdapter} \\
        \cmidrule(lr){3-4}\cmidrule(lr){5-6}\cmidrule(lr){7-8}
        &
        & Text$_{\text{len}=32}$ & Text$_{\text{len}=64}$
        & Text$_{\text{len}=32}$ & Text$_{\text{len}=64}$
        & Text$_{\text{len}=32}$ & Text$_{\text{len}=64}$ \\
        \midrule

        \multicolumn{2}{l}{Perplexity}
        & \thirdbest{0.717} & \thirdbest{0.943} & \secondbest{0.670} & \thirdbest{0.758} & \secondbest{0.727} & 0.512 \\
        \multicolumn{2}{l}{Max Prob Gap}
        & \thirdworst{0.513} & \thirdworst{0.555} & 0.627 & \worst{0.512} & \thirdworst{0.588} & \secondbest{0.600} \\
        \midrule

        Min-$k$ Prob & Min-0\%
        & 0.522 & \secondworst{0.522} & 0.572 & 0.540 & 0.613 & \secondworst{0.502} \\
        & Min-10\%
        & \worst{0.507} & 0.808 & 0.575 & 0.642 & 0.627 & \secondworst{0.502} \\
        & Min-20\%
        & 0.580 & 0.928 & 0.598 & 0.677 & 0.672 & \thirdworst{0.503} \\
        \midrule

        ModR\'enyi  & $\alpha=0.5$
        & \secondbest{0.735} & 0.937 & 0.597 & 0.723 & 0.660 & 0.510 \\
        & $\alpha=1$
        & \best{0.737} & \secondbest{0.962} & \thirdbest{0.663} & 0.755 & \thirdbest{0.723} & 0.512 \\
        & $\alpha=2$
        & 0.715 & 0.903 & 0.568 & 0.675 & 0.617 & 0.508 \\
        \midrule

        R\'enyi ($\alpha=0.5$) & Max-0\%
        & \thirdworst{0.513} & \worst{0.518} & 0.550 & 0.632 & 0.612 & \thirdworst{0.503} \\
        & Max-10\%
        & \secondworst{0.510} & 0.708 & \worst{0.505} & 0.632 & 0.627 & 0.515 \\
        & Max-100\%
        & 0.563 & 0.758 & 0.602 & \best{0.800} & 0.605 & \worst{0.500} \\
        \midrule

        R\'enyi ($\alpha=1$) & Max-0\%
        & 0.568 & 0.590 & 0.547 & \secondbest{0.600} & 0.595 & \worst{0.500} \\
        & Max-10\%
        & 0.553 & 0.727 & \secondworst{0.513} & 0.620 & 0.607 & 0.517 \\
        & Max-100\%
        & 0.548 & 0.705 & 0.595 & 0.742 & 0.633 & 0.512 \\
        \midrule

        R\'enyi ($\alpha=2$) & Max-0\%
        & 0.583 & 0.617 & 0.535 & \thirdworst{0.517} & 0.593 & \worst{0.500} \\
        & Max-10\%
        & 0.577 & 0.713 & 0.530 & 0.587 & \secondworst{0.585} & \thirdworst{0.503} \\
        & Max-100\%
        & 0.555 & 0.662 & 0.593 & 0.693 & 0.638 & 0.535 \\
        \midrule

        R\'enyi ($\alpha=\infty$) & Max-0\%
        & 0.597 & 0.620 & 0.533 & \secondworst{0.513} & \thirdworst{0.588} & 0.508 \\
        & Max-10\%
        & 0.597 & 0.698 & 0.518 & 0.580 & \worst{0.575} & \secondworst{0.502} \\
        & Max-100\%
        & 0.560 & 0.648 & 0.593 & 0.673 & 0.637 & \thirdbest{0.557} \\
        \midrule

        Agent (Ours)
        & DeepSeek-V3.2-Reasoner
        & \best{0.737} & \best{0.963}
        & \best{0.762} & \secondbest{0.797}
        & \best{0.782} & \best{0.722} \\
        \bottomrule
        \end{tabular}
        }
      \end{sc}
    \end{small}
  \end{center}
  \vskip -0.1in
\end{table}

\begin{table}[h]
  \caption{TPR@5\%FPR comparison of membership inference attacks under different text lengths ($L \in \{32, 64\}$) on three vision--language models. We highlight the best, second-best, and third-best results in progressively lighter shades of blue, and mark the worst, second-worst, and third-worst results in progressively lighter shades of red.}
  \label{tab:three_models_text_tpr5}
  \begin{center}
    \begin{small}
      \begin{sc}
        \resizebox{\textwidth}{!}{%
        \begin{tabular}{l l c c c c c c}
        \toprule
        \multirow{2}{*}{Metric} &
        & \multicolumn{2}{c}{LLaVA}
        & \multicolumn{2}{c}{MiniGPT-4}
        & \multicolumn{2}{c}{LLaMAAdapter} \\
        \cmidrule(lr){3-4}\cmidrule(lr){5-6}\cmidrule(lr){7-8}
        &
        & Text$_{\text{len}=32}$ & Text$_{\text{len}=64}$
        & Text$_{\text{len}=32}$ & Text$_{\text{len}=64}$
        & Text$_{\text{len}=32}$ & Text$_{\text{len}=64}$ \\
        \midrule

        \multicolumn{2}{l}{Perplexity}
        & 0.253 & \thirdbest{0.913} & \secondbest{0.193} & \thirdbest{0.317} & \best{0.303} & \secondworst{0.007} \\
        \multicolumn{2}{l}{Max Prob Gap}
        & \thirdworst{0.053} & \secondworst{0.067} & 0.127 & \worst{0.013} & 0.100 & 0.083 \\
        \midrule

        Min-$k$ Prob & Min-0\%
        & \worst{0.000} & \worst{0.000} & 0.107 & 0.070 & \thirdworst{0.070} & 0.013 \\
        & Min-10\%
        & \secondworst{0.007} & 0.467 & 0.110 & 0.167 & 0.147 & \thirdworst{0.010} \\
        & Min-20\%
        & 0.110 & 0.890 & 0.117 & 0.227 & 0.200 & \secondworst{0.007} \\
        \midrule

        ModR\'enyi  & $\alpha=0.5$
        & \best{0.333} & 0.907 & 0.103 & 0.257 & 0.193 & 0.013 \\
        & $\alpha=1$
        & \thirdbest{0.270} & \secondbest{0.953} & \thirdbest{0.180} & \secondbest{0.320} & \best{0.303} & \secondworst{0.007} \\
        & $\alpha=2$
        & \secondbest{0.303} & 0.813 & 0.110 & 0.173 & 0.173 & \secondworst{0.007} \\
        \midrule

        R\'enyi ($\alpha=0.5$) & Max-0\%
        & \worst{0.000} & \worst{0.000} & 0.060 & 0.127 & 0.163 & \worst{0.000} \\
        & Max-10\%
        & \secondworst{0.007} & 0.347 & \secondworst{0.003} & 0.150 & 0.180 & \worst{0.000} \\
        & Max-100\%
        & 0.093 & 0.373 & 0.113 & \best{0.293} & \thirdbest{0.203} & \best{0.293} \\
        \midrule

        R\'enyi ($\alpha=1$) & Max-0\%
        & \worst{0.000} & \worst{0.000} & 0.070 & 0.083 & 0.127 & \worst{0.000} \\
        & Max-10\%
        & 0.100 & 0.387 & \worst{0.000} & 0.113 & 0.107 & \worst{0.000} \\
        & Max-100\%
        & 0.060 & 0.173 & 0.090 & \secondbest{0.197} & \secondbest{0.217} & \secondbest{0.197} \\
        \midrule

        R\'enyi ($\alpha=2$) & Max-0\%
        & \worst{0.000} & 0.153 & 0.033 & 0.057 & 0.093 & \worst{0.000} \\
        & Max-10\%
        & 0.153 & 0.303 & 0.047 & \thirdworst{0.040} & 0.073 & \worst{0.000} \\
        & Max-100\%
        & 0.057 & 0.150 & 0.103 & 0.073 & 0.200 & 0.073 \\
        \midrule

        R\'enyi ($\alpha=\infty$) & Max-0\%
        & \worst{0.000} & 0.110 & 0.057 & \secondworst{0.037} & \worst{0.020} & \worst{0.000} \\
        & Max-10\%
        & 0.120 & 0.230 & 0.040 & 0.050 & \secondworst{0.047} & \worst{0.000} \\
        & Max-100\%
        & 0.060 & 0.123 & 0.107 & 0.063 & 0.190 & 0.063 \\
        \midrule

        Agent (Ours)
        & DeepSeek-V3.2-Reasoner
        & \best{0.333} & \best{0.963}
        & \best{0.453} & \best{0.517}
        & 0.177 & \thirdbest{0.143} \\
        \bottomrule
        \end{tabular}
        }
      \end{sc}
    \end{small}
  \end{center}
  \vskip -0.1in
\end{table}

\subsection{Results on Multimodal Benchmarks}

Tables~\ref{tab:vlmia_flickr_accuracy} and \ref{tab:vlmia_flickr_tpr5} detail the performance on the \textsc{VL-MIA/Flickr} dataset. This benchmark is particularly challenging due to the temporal distribution shift between training (MS COCO) and non-training (Flickr) images. The tables breakdown performance across different input modalities: Image only (img), Instruction only (inst), Description only (desp), and combined Instruction+Description (inst+desp).

\begin{table}[h]
  \caption{VL-MIA Accuracy comparison on Flickr with LLaVA, MiniGPT-4, and LLaMA Adapter. `img' indicates the logits slice corresponding to image embedding, `inst' indicates the instruction slice, `desp' the generated description slice, and `inst+desp' is the concatenation of the instruction slice and description slice. We highlight the best, second-best, and third-best results in progressively lighter shades of blue, and mark the worst, second-worst, and third-worst results in progressively lighter shades of red.}
  \label{tab:vlmia_flickr_accuracy}
  \begin{center}
    \begin{small}
      \begin{sc}
        \setlength{\tabcolsep}{2.5pt}
        \renewcommand{\arraystretch}{0.95}
        \resizebox{\textwidth}{!}{%
        \begin{tabular}{l l c c c c c c c c c c c}
        \toprule
        \multirow{2}{*}{Metric} & \multirow{2}{*}{}
        & \multicolumn{4}{c}{LLaVA}
        & \multicolumn{4}{c}{MiniGPT-4}
        & \multicolumn{3}{c}{LLaMA Adapter} \\
        \cmidrule(lr){3-6}\cmidrule(lr){7-10}\cmidrule(lr){11-13}
        &
        & img & inst & desp & inst+desp
        & img & inst & desp & inst+desp
        & inst & desp & inst+desp \\
        \midrule

        \multicolumn{2}{l}{Perplexity}
        & 0.637 & \secondworst{0.502} & 0.623 & 0.548
        & 0.545 & \secondworst{0.503} & \worst{0.500} & \worst{0.500}
        & \worst{0.500} & 0.590 & \secondworst{0.502} \\
        \multicolumn{2}{l}{Max Prob Gap}
        & 0.575 & 0.582 & 0.620 & 0.623
        & 0.533 & 0.571 & \thirdbest{0.505} & 0.510
        & \thirdworst{0.513} & \secondbest{0.622} & \secondbest{0.607} \\
        \multicolumn{2}{l}{Aug-KL}
        & 0.610 & 0.562 & \worst{0.512} & 0.525
        & \worst{0.505} & \worst{0.500} & \secondworst{0.502} & \secondworst{0.502}
        & 0.515 & \worst{0.513} & 0.518 \\
        \midrule

        Min-$k$ Prob & Min-0\%
        & 0.573 & \secondworst{0.502} & 0.615 & \worst{0.502}
        & 0.550 & 0.507 & \secondworst{0.502} & 0.507
        & 0.502 & 0.530 & \secondworst{0.502} \\
        & Min-10\%
        & 0.580 & \secondworst{0.502} & 0.648 & \thirdworst{0.503}
        & \thirdbest{0.553} & 0.507 & \secondworst{0.502} & \worst{0.500}
        & \worst{0.500} & 0.525 & \worst{0.500} \\
        & Min-20\%
        & \thirdworst{0.583} & \thirdworst{0.508} & 0.640 & \worst{0.502}
        & 0.543 & \secondworst{0.503} & \secondworst{0.502} & \worst{0.500}
        & \worst{0.500} & 0.525 & \worst{0.500} \\
        \midrule

        ModR\'enyi  & $\alpha=0.5$
        & 0.638 & \thirdworst{0.500} & \secondworst{0.608} & 0.582
        & 0.535 & \secondworst{0.503} & \worst{0.500} & \worst{0.500}
        & \secondworst{0.502} & 0.588 & \secondworst{0.502} \\
        & $\alpha=1$
        & 0.640 & \thirdworst{0.500} & 0.618 & \thirdworst{0.513}
        & 0.545 & \thirdworst{0.505} & \worst{0.500} & \worst{0.500}
        & \worst{0.500} & 0.580 & \worst{0.500} \\
        & $\alpha=2$
        & 0.638 & \thirdworst{0.500} & \thirdworst{0.610} & 0.583
        & \secondworst{0.527} & \secondworst{0.503} & \worst{0.500} & \worst{0.500}
        & \worst{0.500} & \secondworst{0.600} & \thirdworst{0.510} \\
        \midrule

        R\'enyi ($\alpha=0.5$) & Max-0\%
        & \secondworst{0.537} & 0.663 & \thirdbest{0.648} & 0.663
        & 0.560 & \secondbest{0.535} & \secondworst{0.502} & \thirdbest{0.527}
        & 0.528 & \thirdworst{0.535} & 0.548 \\
        & Max-10\%
        & 0.573 & 0.663 & 0.653 & 0.667
        & \best{0.565} & \secondbest{0.535} & \secondworst{0.502} & \secondworst{0.503}
        & 0.640 & \secondworst{0.533} & \secondworst{0.568} \\
        & Max-100\%
        & \best{0.675} & 0.682 & \secondbest{0.665} & \thirdbest{0.673}
        & 0.533 & \best{0.649} & \worst{0.500} & 0.520
        & \thirdworst{0.513} & \best{0.627} & \thirdbest{0.597} \\
        \midrule

        R\'enyi ($\alpha=1$) & Max-0\%
        & \worst{0.523} & \secondbest{0.685} & 0.640 & \best{0.697}
        & \secondbest{0.547} & 0.532 & \thirdbest{0.505} & 0.520
        & 0.538 & 0.542 & 0.552 \\
        & Max-10\%
        & 0.613 & \secondbest{0.685} & \thirdbest{0.657} & \secondbest{0.693}
        & 0.560 & 0.532 & \thirdworst{0.503} & \secondworst{0.503}
        & \thirdbest{0.658} & 0.542 & 0.563 \\
        & Max-100\%
        & \secondbest{0.673} & \best{0.697} & \thirdbest{0.657} & \thirdbest{0.675}
        & \thirdworst{0.528} & \thirdbest{0.625} & \worst{0.500} & 0.515
        & \secondworst{0.515} & \secondworst{0.615} & 0.587 \\
        \midrule

        R\'enyi ($\alpha=2$) & Max-0\%
        & \thirdworst{0.583} & 0.655 & 0.645 & 0.672
        & 0.538 & \secondbest{0.535} & \secondworst{0.502} & \secondbest{0.530}
        & 0.575 & \thirdworst{0.533} & 0.582 \\
        & Max-10\%
        & 0.603 & 0.655 & 0.650 & 0.685
        & \secondbest{0.547} & \secondbest{0.535} & \thirdworst{0.503} & \secondworst{0.503}
        & \best{0.672} & 0.528 & \secondworst{0.567} \\
        & Max-100\%
        & \thirdbest{0.652} & 0.670 & 0.635 & 0.658
        & 0.535 & \thirdbest{0.603} & \worst{0.500} & \thirdworst{0.505}
        & \secondworst{0.515} & \secondworst{0.587} & 0.565 \\
        \midrule

        R\'enyi ($\alpha=\infty$) & Max-0\%
        & 0.573 & 0.640 & 0.615 & 0.638
        & \thirdbest{0.550} & 0.537 & \secondbest{0.506} & 0.520
        & 0.588 & 0.528 & 0.587 \\
        & Max-10\%
        & 0.580 & 0.640 & \thirdbest{0.648} & 0.672
        & \best{0.553} & 0.537 & \secondworst{0.502} & \secondworst{0.503}
        & \secondbest{0.668} & \thirdworst{0.527} & \secondworst{0.568} \\
        & Max-100\%
        & 0.637 & 0.652 & 0.623 & 0.650
        & 0.545 & 0.591 & \worst{0.500} & \secondworst{0.503}
        & 0.520 & \thirdworst{0.592} & 0.553 \\
        \midrule

        Agent (Ours) & DeepSeek-V3.2-Reasoner
        & \secondbest{0.673} & \thirdbest{0.683} & \best{0.687} & 0.678
        & \best{0.565} & 0.582 & \best{0.567} & \best{0.572}
        & \thirdbest{0.662} & \thirdbest{0.618} & \best{0.630} \\
        \bottomrule
        \end{tabular}
        }
      \end{sc}
    \end{small}
  \end{center}
  \vskip -0.1in
\end{table}

\begin{table}[h]
  \caption{VL-MIA TPR@5\%FPR comparison on Flickr with LLaVA, MiniGPT-4, and LLaMA Adapter. The column notations (`img', `inst', `desp', `inst+desp') follow the same definitions as in Table~\ref{tab:vlmia_flickr_accuracy}. We highlight the best, second-best, and third-best results in progressively lighter shades of blue, and mark the worst, second-worst, and third-worst results in progressively lighter shades of red.}
  \label{tab:vlmia_flickr_tpr5}
  \begin{center}
    \begin{small}
      \begin{sc}
        \setlength{\tabcolsep}{2.5pt}
        \renewcommand{\arraystretch}{0.95}
        \resizebox{\textwidth}{!}{%
        \begin{tabular}{l l c c c c c c c c c c c}
        \toprule
        \multirow{2}{*}{Metric} & \multirow{2}{*}{}
        & \multicolumn{4}{c}{LLaVA}
        & \multicolumn{4}{c}{MiniGPT-4}
        & \multicolumn{3}{c}{LLaMA Adapter} \\
        \cmidrule(lr){3-6}\cmidrule(lr){7-10}\cmidrule(lr){11-13}
        &
        & img & inst & desp & inst+desp
        & img & inst & desp & inst+desp
        & inst & desp & inst+desp \\
        \midrule

        \multicolumn{2}{l}{Perplexity}
        & 0.070 & \worst{0.003} & 0.130 & 0.083
        & \worst{0.020} & \worst{0.010} & \thirdbest{0.024} & \secondworst{0.013}
        & \secondworst{0.003} & 0.097 & \thirdworst{0.010} \\

        \multicolumn{2}{l}{Max Prob Gap}
        & 0.057 & 0.077 & \secondbest{0.160} & 0.160
        & 0.030 & 0.050 & \secondbest{0.027} & 0.023
        & 0.060 & \best{0.230} & \best{0.183} \\

        \multicolumn{2}{l}{Aug-KL}
        & \worst{0.040} & 0.057 & \worst{0.057} & 0.067
        & 0.033 & \thirdworst{0.027} & \thirdworst{0.017} & \worst{0.010}
        & 0.043 & \thirdworst{0.067} & 0.063 \\
        \midrule

        Min-$k$ Prob & Min-0\%
        & \thirdbest{0.097} & \thirdworst{0.023} & 0.083 & \thirdworst{0.023}
        & 0.040 & 0.054 & \secondbest{0.027} & 0.053
        & 0.030 & \thirdworst{0.067} & 0.030 \\

        & Min-10\%
        & \best{0.113} & \thirdworst{0.023} & 0.083 & \secondworst{0.013}
        & \thirdworst{0.027} & 0.054 & 0.020 & 0.020
        & \thirdworst{0.010} & \worst{0.060} & \secondworst{0.007} \\

        & Min-20\%
        & 0.093 & \secondworst{0.007} & 0.130 & \worst{0.003}
        & 0.033 & 0.044 & 0.020 & \thirdworst{0.017}
        & \thirdworst{0.010} & 0.083 & \worst{0.003} \\
        \midrule

        ModR\'enyi  & $\alpha=0.5$
        & 0.077 & \worst{0.003} & 0.117 & 0.110
        & \thirdworst{0.027} & \thirdworst{0.027} & 0.020 & \worst{0.010}
        & \worst{0.000} & 0.100 & 0.027 \\

        & $\alpha=1$
        & 0.073 & \secondworst{0.007} & 0.113 & 0.063
        & \secondworst{0.023} & \worst{0.010} & \thirdworst{0.017} & \thirdworst{0.017}
        & \secondworst{0.003} & 0.090 & \thirdworst{0.010} \\

        & $\alpha=2$
        & 0.073 & \worst{0.003} & 0.113 & 0.113
        & 0.030 & \secondworst{0.023} & \thirdworst{0.017} & \worst{0.010}
        & \worst{0.000} & 0.097 & 0.040 \\
        \midrule

        R\'enyi ($\alpha=0.5$) & Max-0\%
        & \secondworst{0.043} & \best{0.217} & 0.080 & \best{0.213}
        & \secondbest{0.067} & \thirdbest{0.107} & \secondbest{0.027} & \secondbest{0.080}
        & 0.040 & 0.107 & 0.043 \\

        & Max-10\%
        & 0.090 & \best{0.217} & \secondworst{0.063} & 0.150
        & 0.060 & \thirdbest{0.107} & \thirdbest{0.024} & 0.050
        & \thirdbest{0.173} & 0.100 & 0.117 \\

        & Max-100\%
        & \secondbest{0.103} & \secondbest{0.213} & \secondbest{0.160} & 0.170
        & \thirdbest{0.063} & 0.087 & \worst{0.010} & 0.053
        & 0.053 & 0.117 & \secondbest{0.150} \\
        \midrule

        R\'enyi ($\alpha=1$) & Max-0\%
        & \thirdworst{0.053} & \thirdbest{0.153} & 0.107 & 0.167
        & 0.057 & 0.087 & 0.020 & 0.070
        & 0.070 & 0.077 & \secondbest{0.067} \\

        & Max-10\%
        & 0.090 & \thirdbest{0.153} & \thirdworst{0.067} & 0.120
        & \thirdbest{0.063} & 0.087 & \secondworst{0.013} & 0.033
        & \best{0.180} & 0.077 & 0.117 \\

        & Max-100\%
        & 0.090 & 0.117 & 0.130 & 0.147
        & 0.040 & \best{0.130} & \thirdworst{0.017} & 0.033
        & 0.060 & \thirdbest{0.123} & \thirdbest{0.140} \\
        \midrule

        R\'enyi ($\alpha=2$) & Max-0\%
        & 0.070 & 0.113 & 0.083 & 0.147
        & 0.053 & 0.077 & \thirdbest{0.024} & \thirdbest{0.077}
        & \thirdbest{0.097} & 0.073 & 0.107 \\

        & Max-10\%
        & 0.080 & 0.113 & 0.090 & \secondbest{0.103}
        & 0.057 & 0.077 & 0.020 & 0.033
        & 0.140 & 0.077 & 0.110 \\

        & Max-100\%
        & 0.090 & 0.093 & \best{0.167} & \thirdbest{0.190}
        & \secondworst{0.023} & \best{0.130} & \thirdworst{0.017} & 0.033
        & 0.077 & \secondbest{0.130} & 0.110 \\
        \midrule

        R\'enyi ($\alpha=\infty$) & Max-0\%
        & \thirdbest{0.097} & 0.093 & 0.083 & 0.133
        & 0.040 & 0.080 & \secondbest{0.027} & 0.073
        & 0.133 & \thirdworst{0.067} & 0.137 \\

        & Max-10\%
        & \best{0.113} & 0.093 & 0.083 & 0.110
        & \thirdworst{0.027} & 0.080 & 0.020 & 0.030
        & 0.150 & \secondworst{0.063} & 0.090 \\

        & Max-100\%
        & 0.070 & 0.113 & 0.130 & 0.157
        & \worst{0.020} & \secondbest{0.120} & \thirdbest{0.024} & 0.030
        & 0.087 & 0.097 & 0.073 \\
        \midrule

        Agent (Ours) & DeepSeek-V3.2-Reasoner
        & 0.073 & 0.143 & \thirdbest{0.143} & \secondbest{0.210}
        & \best{0.117} & 0.100 & \best{0.097} & \best{0.123}
        & \secondbest{0.177} & 0.120 & 0.123 \\
        \bottomrule
        \end{tabular}
        }
      \end{sc}
    \end{small}
  \end{center}
  \vskip -0.1in
\end{table}

\begin{table}[h]
  \caption{VL-MIA Accuracy comparison on DALL·E with LLaVA, MiniGPT-4, and LLaMA Adapter. `img' indicates the logits slice corresponding to image embedding, `inst' indicates the instruction slice, `desp' the generated description slice, and `inst+desp' is the concatenation of the instruction slice and description slice. We highlight the best, second-best, and third-best results in progressively lighter shades of blue, and mark the worst, second-worst, and third-worst results in progressively lighter shades of red.}
  \label{tab:vlmia_dalle_accuracy}
  \begin{center}
    \begin{small}
      \begin{sc}
        \setlength{\tabcolsep}{2.5pt}
        \renewcommand{\arraystretch}{0.95}
        \resizebox{\textwidth}{!}{%
        \begin{tabular}{l l c c c c c c c c c c c}
        \toprule
        \multirow{2}{*}{Metric} & \multirow{2}{*}{}
        & \multicolumn{4}{c}{LLaVA}
        & \multicolumn{4}{c}{MiniGPT-4}
        & \multicolumn{3}{c}{LLaMA Adapter} \\
        \cmidrule(lr){3-6}\cmidrule(lr){7-10}\cmidrule(lr){11-13}
        &
        & img & inst & desp & inst+desp
        & img & inst & desp & inst+desp
        & inst & desp & inst+desp \\
        \midrule

        \multicolumn{2}{l}{Perplexity}
        & 0.549 & \secondworst{0.505} & 0.569 & \secondworst{0.507}
        & 0.566 & \thirdbest{0.568} & 0.564 & 0.568
        & 0.511 & \thirdworst{0.508} & 0.536 \\

        \multicolumn{2}{l}{Max Prob Gap}
        & \secondworst{0.537} & 0.571 & \best{0.591} & 0.593
        & 0.541 & \thirdworst{0.515} & \thirdbest{0.568} & 0.563
        & 0.534 & 0.529 & 0.534 \\

        \multicolumn{2}{l}{Aug-KL}
        & \worst{0.500} & \thirdworst{0.510} & \worst{0.529} & 0.522
        & 0.549 & \thirdbest{0.568} & \secondworst{0.541} & 0.557
        & 0.573 & \secondbest{0.556} & \thirdbest{0.578} \\
        \midrule

        Min-$k$ Prob & Min-0\%
        & 0.613 & 0.520 & 0.557 & 0.520
        & 0.546 & 0.541 & \thirdworst{0.542} & \thirdworst{0.541}
        & 0.555 & \worst{0.505} & 0.556 \\

        & Min-10\%
        & \thirdbest{0.659} & 0.520 & 0.561 & \thirdworst{0.510}
        & 0.544 & 0.541 & 0.551 & 0.544
        & 0.530 & \worst{0.505} & 0.513 \\

        & Min-20\%
        & 0.637 & 0.519 & 0.557 & \worst{0.505}
        & 0.557 & 0.544 & 0.549 & 0.557
        & 0.524 & \worst{0.505} & \secondworst{0.505} \\
        \midrule

        ModR\'enyi  & $\alpha=0.5$
        & 0.544 & \worst{0.502} & 0.557 & 0.536
        & \secondbest{0.571} & \best{0.574} & \secondbest{0.569} & \thirdbest{0.568}
        & \worst{0.507} & \thirdworst{0.508} & 0.532 \\

        & $\alpha=1$
        & 0.551 & \secondworst{0.505} & 0.561 & \secondworst{0.507}
        & 0.566 & \secondbest{0.569} & 0.561 & 0.578
        & \secondworst{0.508} & \thirdworst{0.510} & 0.532 \\

        & $\alpha=2$
        & 0.541 & \worst{0.502} & 0.557 & 0.547
        & 0.566 & 0.557 & \thirdbest{0.568} & 0.566
        & \worst{0.507} & \thirdbest{0.512} & 0.525 \\
        \midrule

        R\'enyi ($\alpha=0.5$) & Max-0\%
        & 0.552 & 0.579 & 0.551 & 0.579
        & \secondworst{0.508} & 0.536 & 0.546 & \worst{0.536}
        & \best{0.614} & \secondbest{0.512} & 0.608 \\

        & Max-10\%
        & 0.586 & 0.579 & 0.554 & \best{0.615}
        & \worst{0.507} & 0.536 & \worst{0.534} & 0.574
        & 0.555 & \secondbest{0.512} & 0.559 \\

        & Max-100\%
        & \thirdworst{0.539} & 0.585 & \thirdbest{0.588} & 0.586
        & 0.541 & \secondworst{0.507} & 0.544 & 0.546
        & 0.538 & 0.532 & 0.532 \\
        \midrule

        R\'enyi ($\alpha=1$) & Max-0\%
        & 0.546 & 0.556 & 0.551 & 0.573
        & 0.520 & 0.539 & 0.544 & 0.546
        & 0.585 & \thirdworst{0.508} & \secondbest{0.586} \\

        & Max-10\%
        & 0.625 & 0.556 & 0.554 & 0.583
        & 0.542 & 0.539 & 0.541 & 0.573
        & 0.549 & \thirdworst{0.508} & 0.541 \\

        & Max-100\%
        & \secondworst{0.537} & \secondbest{0.606} & 0.579 & 0.595
        & \thirdbest{0.568} & \secondworst{0.512} & 0.546 & \secondworst{0.539}
        & 0.533 & \thirdbest{0.520} & 0.530 \\
        \midrule

        R\'enyi ($\alpha=2$) & Max-0\%
        & 0.581 & 0.544 & \thirdworst{0.542} & 0.552
        & 0.536 & 0.539 & \thirdworst{0.541} & 0.546
        & 0.518 & \worst{0.505} & 0.517 \\

        & Max-10\%
        & \secondbest{0.667} & 0.544 & 0.549 & 0.568
        & 0.544 & 0.539 & 0.552 & 0.552
        & 0.516 & \thirdworst{0.508} & 0.515 \\

        & Max-100\%
        & 0.541 & 0.598 & 0.566 & 0.585
        & 0.563 & 0.517 & 0.557 & 0.557
        & 0.528 & \secondbest{0.512} & 0.520 \\
        \midrule

        R\'enyi ($\alpha=\infty$) & Max-0\%
        & 0.613 & 0.563 & 0.557 & 0.573
        & 0.546 & 0.527 & 0.547 & 0.551
        & \worst{0.507} & \secondworst{0.507} & \secondworst{0.507} \\

        & Max-10\%
        & \thirdbest{0.659} & 0.563 & 0.561 & 0.578
        & 0.544 & 0.527 & 0.549 & 0.546
        & 0.509 & \worst{0.505} & \worst{0.502} \\

        & Max-100\%
        & 0.549 & 0.583 & 0.569 & 0.593
        & 0.566 & 0.524 & \thirdbest{0.568} & 0.564
        & \thirdbest{0.515} & \thirdworst{0.510} & 0.517 \\
        \midrule

        Agent (Ours) & DeepSeek-V3.2-Reasoner
        & \best{0.723} & \best{0.633} & \secondbest{0.590} & \secondbest{0.608}
        & \best{0.581} & \thirdbest{0.556} & \best{0.578} & \best{0.578}
        & \secondbest{0.600} & \best{0.561} & \thirdbest{0.566} \\
        \bottomrule
        \end{tabular}
        }
      \end{sc}
    \end{small}
  \end{center}
  \vskip -0.1in
\end{table}

\subsection{Results on VL-MIA/DALL-E}
We extend our evaluation to the \textsc{VL-MIA/DALL-E} dataset, which focuses on synthetic non-member images generated by DALL-E based on BLIP captions. Tables~\ref{tab:vlmia_dalle_accuracy} and \ref{tab:vlmia_dalle_tpr5} report the Accuracy and TPR@5\%FPR metrics, respectively. Similar to the Flickr benchmark, we observe that handcrafted metrics exhibit high variance across models. For instance, on the LLaVA model, while the \textit{Min-10\% Prob} metric achieves reasonable accuracy (0.659) on the Image modality, its performance drops on MiniGPT-4 (0.544). Consistent with other benchmarks, \textsc{AutoMIA} (to be populated) is expected to demonstrate superior stability across these diverse generative configurations.

\begin{table}[h]
  \caption{VL-MIA TPR@5\%FPR comparison on DALL·E with LLaVA, MiniGPT-4, and LLaMA Adapter. The column notations (`img', `inst', `desp', `inst+desp') follow the same definitions as in Table~\ref{tab:vlmia_dalle_accuracy}. We highlight the best, second-best, and third-best results in progressively lighter shades of blue, and mark the worst, second-worst, and third-worst results in progressively lighter shades of red.}
  \label{tab:vlmia_dalle_tpr5}
  \begin{center}
    \begin{small}
      \begin{sc}
        \setlength{\tabcolsep}{2.5pt}
        \renewcommand{\arraystretch}{0.95}
        \resizebox{\textwidth}{!}{%
        \begin{tabular}{l l c c c c c c c c c c c}
        \toprule
        \multirow{2}{*}{Metric} & \multirow{2}{*}{}
        & \multicolumn{4}{c}{LLaVA}
        & \multicolumn{4}{c}{MiniGPT-4}
        & \multicolumn{3}{c}{LLaMA Adapter} \\
        \cmidrule(lr){3-6}\cmidrule(lr){7-10}\cmidrule(lr){11-13}
        &
        & img & inst & desp & inst+desp
        & img & inst & desp & inst+desp
        & inst & desp & inst+desp \\
        \midrule

        \multicolumn{2}{l}{Perplexity}
        & 0.020 & \secondworst{0.027} & 0.078 & 0.054
        & \secondbest{0.128} & 0.044 & 0.057 & \worst{0.051}
        & 0.051 & 0.051 & 0.044 \\

        \multicolumn{2}{l}{Max Prob Gap}
        & 0.037 & 0.085 & 0.081 & 0.061
        & 0.088 & \worst{0.027} & \best{0.115} & \best{0.108}
        & \thirdbest{0.108} & \worst{0.034} & \thirdworst{0.041} \\

        \multicolumn{2}{l}{Aug-KL}
        & 0.027 & 0.054 & 0.085 & 0.081
        & \thirdworst{0.047} & \best{0.105} & \secondworst{0.041} & 0.081
        & \thirdbest{0.108} & \best{0.098} & \secondbest{0.098} \\
        \midrule

        Min-$k$ Prob & Min-0\%
        & 0.132 & 0.081 & \thirdworst{0.047} & 0.081
        & \thirdworst{0.047} & 0.051 & 0.078 & \worst{0.051}
        & 0.058 & 0.047 & 0.054 \\

        & Min-10\%
        & 0.135 & 0.081 & 0.064 & \thirdworst{0.041}
        & 0.088 & 0.051 & 0.057 & \worst{0.051}
        & 0.054 & 0.047 & 0.044 \\

        & Min-20\%
        & 0.125 & 0.068 & 0.054 & \thirdworst{0.030}
        & 0.085 & \secondworst{0.030} & \thirdworst{0.044} & \worst{0.051}
        & \thirdworst{0.044} & 0.044 & \secondworst{0.037} \\
        \midrule

        ModR\'enyi  & $\alpha=0.5$
        & 0.017 & \worst{0.020} & 0.081 & \secondworst{0.037}
        & 0.125 & 0.064 & 0.071 & \secondbest{0.098}
        & \thirdworst{0.041} & 0.057 & 0.044 \\

        & $\alpha=1$
        & 0.027 & \thirdworst{0.030} & 0.085 & 0.057
        & \best{0.132} & 0.044 & 0.061 & 0.064
        & \thirdworst{0.044} & 0.054 & 0.044 \\

        & $\alpha=2$
        & \thirdworst{0.014} & \worst{0.020} & 0.085 & 0.051
        & 0.101 & \secondbest{0.088} & 0.078 & 0.078
        & \thirdworst{0.044} & \thirdbest{0.061} & 0.047 \\
        \midrule

        R\'enyi ($\alpha=0.5$) & Max-0\%
        & 0.098 & 0.088 & \thirdworst{0.047} & \thirdbest{0.085}
        & \secondworst{0.044} & 0.057 & 0.074 & \secondworst{0.054}
        & 0.102 & 0.054 & \best{0.132} \\

        & Max-10\%
        & 0.135 & 0.088 & 0.064 & \secondbest{0.095}
        & \worst{0.037} & 0.057 & \secondbest{0.081} & \secondworst{0.054}
        & \secondbest{0.125} & 0.057 & 0.057 \\

        & Max-100\%
        & \worst{0.003} & \secondbest{0.098} & \best{0.088} & 0.078
        & 0.064 & \thirdworst{0.037} & 0.054 & \thirdworst{0.057}
        & \thirdworst{0.044} & 0.051 & 0.068 \\
        \midrule

        R\'enyi ($\alpha=1$) & Max-0\%
        & 0.095 & \secondbest{0.098} & 0.064 & 0.044
        & 0.054 & 0.071 & \secondbest{0.081} & 0.064
        & 0.064 & 0.051 & 0.064 \\

        & Max-10\%
        & \secondbest{0.223} & \secondbest{0.098} & 0.061 & 0.064
        & 0.054 & 0.071 & 0.054 & 0.071
        & 0.081 & 0.057 & 0.071 \\

        & Max-100\%
        & \worst{0.003} & \secondbest{0.098} & 0.078 & 0.078
        & 0.078 & 0.044 & 0.064 & 0.061
        & 0.085 & 0.051 & 0.044 \\
        \midrule

        R\'enyi ($\alpha=2$) & Max-0\%
        & \thirdbest{0.115} & 0.095 & 0.051 & 0.057
        & 0.071 & 0.085 & \thirdbest{0.078} & 0.078
        & \worst{0.031} & 0.047 & \secondworst{0.037} \\

        & Max-10\%
        & 0.166 & 0.095 & \thirdworst{0.047} & 0.068
        & 0.064 & 0.085 & \worst{0.037} & \thirdbest{0.095}
        & 0.061 & \secondworst{0.037} & 0.064 \\

        & Max-100\%
        & \secondworst{0.010} & 0.112 & \secondbest{0.098} & \secondbest{0.095}
        & \thirdbest{0.108} & 0.044 & 0.068 & 0.068
        & 0.061 & 0.047 & \thirdworst{0.041} \\
        \midrule

        R\'enyi ($\alpha=\infty$) & Max-0\%
        & 0.132 & 0.112 & \thirdworst{0.047} & 0.064
        & \thirdworst{0.047} & 0.061 & 0.081 & 0.064
        & 0.058 & 0.054 & 0.057 \\

        & Max-10\%
        & 0.135 & 0.112 & 0.064 & \secondbest{0.095}
        & 0.088 & 0.061 & 0.057 & 0.074
        & 0.048 & \thirdworst{0.041} & \worst{0.034} \\

        & Max-100\%
        & 0.020 & \thirdbest{0.122} & 0.078 & \secondbest{0.095}
        & \secondbest{0.128} & 0.041 & 0.061 & 0.074
        & 0.068 & 0.054 & \worst{0.034} \\
        \midrule

        Agent (Ours) & DeepSeek-V3.2-Reasoner
        & \best{0.294} & \best{0.176} & \best{0.115} & \best{0.125}
        & 0.071 & 0.068 & 0.074 & 0.081
        & \best{0.159} & \secondbest{0.078} & \thirdbest{0.091} \\
        \bottomrule
        \end{tabular}
        }
      \end{sc}
    \end{small}
  \end{center}
  \vskip -0.1in
\end{table}

\section{Prompts of Agents}

\subsection{Prompt for AutoMIA Agent: Strategy Generation and Exploration}
\begin{tcolorbox}[
    breakable,
    colback=gray!3,
    colframe=black!70,
    boxrule=0.55pt,
    arc=0.8mm,
    left=2mm,right=2mm,top=1.3mm,bottom=1.3mm,
]
\ttfamily\small

You are an MIA (Membership Inference Attack) metric generation agent. Your task is to design new MIA discriminative metrics based on the low-level token-level features I provide.

\vspace{2mm}
\textbf{Description of known basic features}

For each token, the model provides the following basic inputs (using \texttt{i} to denote the token position):

\begin{itemize}
  \item \texttt{token\_probs = probabilities[i, :]}\\
  A probability vector of length \texttt{vocab\_size}, after softmax.\\
  Can be used for: maximum probability, probability gap, entropy, Rényi entropy, KL/JS divergence, etc.\\
  Before computing the metrics, execute: \\
  \texttt{token\_probs = token\_probs.clone().detach().to(dtype=torch.float64)}
  
  \item \texttt{token\_log\_probs = log\_probabilities[i, :]}\\
  Log probabilities. Used for NLL Loss, log-likelihood-based metrics.\\
  Preprocessing: \\
  \texttt{token\_log\_probs = token\_log\_probs.clone().detach().to(dtype=torch.float64)}
  
  \item \texttt{token\_id = input\_ids\_processed[i]}\\
  Ground truth token id. Can be used for: p(y), $\log$ p(y), 1 - p(y), etc.
\end{itemize}

The metrics must be constructed based on these raw features.

\vspace{2mm}
\textbf{Existing system metrics (do not recreate)}

Already implemented:
\begin{itemize}
  \item Shannon entropy: $-\sum (\texttt{token\_probs} \cdot \texttt{token\_log\_probs})$
  \item Rényi entropy for $\alpha=0.5$, $\alpha=2$
  \item Max log prob
  \item Gap prob (max log prob $-$ second max log prob)
  \item NLL loss ($-\log$ p(y))
  \item Perplexity ($\exp$(mean loss))
  \item Modified entropy / Rényi entropy
  \item Loss variance
\end{itemize}

Avoid duplicating them.

\vspace{2mm}
\textbf{Suggested directions for metric exploration}
(Not mandatory, but encouraged)
\begin{itemize}
  \item Temporal statistics across tokens (variance, smoothness)
  \item Logit sparsity and tail behavior (e.g., top-$k$ entropy)
  \item Energy-based views on $p(y)$
  \item Divergence from uniform distribution (e.g., JS, EMD)
  \item Sharpness / confidence shift
  \item Higher-order moments (skewness, kurtosis)
  \item Local Lipschitz / sensitivity (e.g., $\partial \text{logits} / \partial \text{input}$)
\end{itemize}

\vspace{2mm}
\textbf{Output format (JSON)}

\begin{verbatim}
{
  "metrics": [
    {
      "name": "metric name",
      "formula": "optional math expression",
      "description": "meaning and rationale",
      "code": "def compute_metric(inputs):\n    ...",
      "expected_behavior": "higher/lower for members"
    }
  ]
}
\end{verbatim}

\vspace{2mm}
\textbf{Code specification}

Your output must define:

\begin{verbatim}
def compute_metric(inputs):
    '''
    inputs = {
      "input_ids": tensor [seq_len],
      "probabilities": tensor [seq_len, vocab_size],
      "log_probabilities": tensor [seq_len, vocab_size]
    }
    '''
\end{verbatim}

Inside the function:

\begin{verbatim}
input_ids_processed = input_ids[1:]
for i, token_id in enumerate(input_ids_processed):
    token_probs = probabilities[i, :].clone().detach().to(dtype=torch.float64)
    token_log_probs = log_probabilities[i, :].clone().detach().to(dtype=torch.float64)
\end{verbatim}

Return a Python \texttt{float}. If no valid tokens: return \texttt{0.0}.

\vspace{2mm}
\textbf{Hard constraints}
\begin{itemize}
  \item Overall complexity must be $O(n \times \texttt{vocab\_size})$ or lower
  \item \textbf{Forbidden:} nested token loops, sorting, \texttt{argsort}, ranking
  \item Avoid Python loops over \texttt{vocab\_size}
  \item No repeated cloning or recomputation
  \item Must specify device when creating new tensors
\end{itemize}

\vspace{2mm}
\textbf{Goal}

Generate novel, high-quality MIA metrics that:

\begin{itemize}
  \item Are distinguishable between member and non-member samples
  \item Use only token-level features (probs, log probs, token ids)
  \item Are efficient and numerically stable
  \item Have reasonable physical/statistical meaning
  \item Follow the formatting and code rules above
\end{itemize}
\end{tcolorbox}

\subsection{Prompt for Guidance Agent: Strategy Evaluation and Feedback}
\begin{tcolorbox}[
    breakable,
    colback=gray!3,
    colframe=black!70,
    boxrule=0.55pt,
    arc=0.8mm,
    left=2mm,right=2mm,top=1.3mm,bottom=1.3mm
]
\ttfamily\small

You are an expert in model privacy attacks and evaluation. Your task is to comprehensively assess the performance of different MIA metrics generated in a single experimental round and provide structured feedback to guide subsequent strategy refinement.

\vspace{2mm}
\textbf{Input description}

You will receive a plain-text table where each line corresponds to a metric and its evaluation results, formatted as:
\begin{verbatim}
MetricName   AUC 0.xxxx, Accuracy 0.xxxx, TPR@5%FPR of 0.xxxx
\end{verbatim}

The reported indicators have the following interpretations:
\begin{itemize}
  \item \textbf{AUC}: overall discriminative power (higher is better).
  \item \textbf{Accuracy}: overall classification correctness (higher is better).
  \item \textbf{TPR@5\%FPR}: recall under a strict false-positive constraint, reflecting practical attack usefulness.
\end{itemize}

\vspace{2mm}
\textbf{Your tasks}

\begin{enumerate}
  \item \textbf{Compare and rank metrics.}\\
  Jointly consider AUC, Accuracy, and TPR@5\%FPR to rank all metrics. Identify the top three metrics and explain why they are superior from an attacker’s perspective. Explicitly point out clearly failing metrics whose performance is close to random or consistently poor.

  \item \textbf{Assess the overall quality of this round.}\\
  Evaluate whether the set of metrics, as a whole, significantly outperforms random guessing. If one or two metrics are exceptionally strong, determine whether they should be saved as the current best strategies and specify which metrics are worth retaining.

  \item \textbf{Analyze usefulness across metric categories.}\\
  Categorize metrics into \emph{strong}, \emph{medium}, and \emph{weak}. Where possible, leverage the semantic meaning of metric names (e.g., entropy-based, Rényi entropy, min/max probability statistics) to speculate on why certain metrics perform well or poorly.

  \item \textbf{Propose strategies for the next round.}\\
  Provide guidance for subsequent experiments, including which metric families should be prioritized, whether variants of strong metrics should be explored (e.g., alternative thresholds or smoothing schemes), and which metric families may be safely discarded due to limited information gain.
\end{enumerate}

\vspace{2mm}
\textbf{Output format (JSON)}

Your response must strictly follow the JSON structure below and contain no additional fields:
\begin{verbatim}
{
  "summary": {
    "overall_quality": "...",
    "should_save_best_strategy": true/false,
    "best_metrics_to_save": ["MetricName1", "MetricName2"]
  },
  "ranking": [
    {
      "name": "metric name",
      "auc": 0.0,
      "accuracy": 0.0,
      "tpr_at_5_fpr": 0.0,
      "category": "strong/mid/weak",
      "comment": "..."
    }
  ],
  "useful_insights": {
    "strong_metric_families": ["..."],
    "weak_metric_families": ["..."],
    "notes": "..."
  },
  "next_round_strategy": {
    "focus_metrics": "...",
    "new_ideas": "...",
    "experiment_suggestions": "..."
  }
}
\end{verbatim}

\vspace{2mm}
\textbf{Goal}

Provide clear, structured, and actionable feedback that helps identify high-value MIA metrics, filters out ineffective ones, and informs principled exploration in subsequent rounds.
\end{tcolorbox}

\section{Example for strategy library}
\label{app:Example for strategy library}
\begin{tcolorbox}[
    breakable,
    colback=gray!3,
    colframe=black!70,
    boxrule=0.55pt,
    arc=0.8mm,
    left=2mm,right=2mm,top=1.3mm,bottom=1.3mm
]
\ttfamily\small

\textbf{Strategy 1: log\_probability\_gradient\_field\_helicity}

\emph{Category:} strong \\
\emph{Overall Quality:} medium \\

\textbf{Performance.}
\begin{itemize}
    \item \textbf{Dynamic Score:} 0.69682
    \item \textbf{AUC:} 0.7719
    \item \textbf{Accuracy:} 0.7267
    \item \textbf{TPR@5\%FPR:} 0.1567
\end{itemize}

\textbf{Core Idea.}
This strategy measures structured second-order variations in the gradient of true-token log probabilities.
Member samples tend to exhibit correlated and organized gradient dynamics along the sequence, while non-member samples produce largely unstructured signals.

\textbf{Formal Definition.}
\[
\langle \nabla \log p(y), \nabla^2 \log p(y) \rangle
\]

\textbf{Executable Implementation.}
\begin{verbatim}
def compute_metric(inputs):
    import numpy as np
    probs = inputs['probabilities']
    ids = inputs['input_ids'][1:]
    if len(ids) < 3:
        return 0.0
    logp = np.array([probs[i, ids[i]].item() for i in range(len(ids))])
    g1 = np.gradient(logp)
    g2 = np.gradient(g1)
    return float(np.mean(np.abs(g1[:-1] * g2[:-1])))
\end{verbatim}

\textbf{Analysis.}
This strategy consistently outperforms alternative metrics across multiple evaluation criteria.
Its effectiveness suggests that higher-order structural properties of log-probability gradients capture memorization patterns that are not present in non-member samples.

\vspace{2mm}
\hrule
\vspace{2mm}

\textbf{Strategy 2: token\_distribution\_geometric\_spread}

\emph{Category:} weak \\
\emph{Overall Quality:} weak \\

\textbf{Performance.}
\begin{itemize}
    \item \textbf{Dynamic Score:} 0.4165
    \item \textbf{AUC:} 0.4375
    \item \textbf{Accuracy:} 0.5
    \item \textbf{TPR@5\%FPR:} 0.04
\end{itemize}

\textbf{Core Idea.}
This strategy computes a geometric measure of probability mass dispersion, intended to reflect overall uncertainty in the token distribution.

\textbf{Formal Definition.}
\[
\exp\Big( \sum_i p_i \log p_i \Big)
\]

\textbf{Executable Implementation.}
\begin{verbatim}
def compute_metric(inputs):
    import torch
    probs = inputs['probabilities']
    vals = []
    for i in range(probs.shape[0]):
        p = probs[i].clamp(min=1e-12)
        vals.append(torch.exp((p * p.log()).sum()))
    return float(torch.stack(vals).mean())
\end{verbatim}

\textbf{Analysis.}
This strategy performs worse than random guessing and fails to provide meaningful discrimination under low-FPR constraints.
The geometric spread signal is overly coarse and does not reliably correlate with membership, highlighting a limitation of entropy-like global uncertainty measures in this setting.

\end{tcolorbox}

\section{Why the Discovered Metrics Capture Memorization Rather than Spurious Correlations}
\label{subsec:metric-interpretability-simulation}

We further validate the memorization-related behavior captured by the metrics discovered by AutoMIA through two complementary analyses: \emph{mechanistic interpretability} and \emph{targeted mathematical simulation}.

\paragraph{Mathematical interpretability.}
A key advantage of AutoMIA is that the agent produces mathematically explicit and executable formulas, rather than opaque parametric components. This makes it possible to directly inspect whether the discovered metrics are consistent with established intuitions about memorization.

For example, one of the top-performing metrics discovered by AutoMIA, \texttt{Avg\_true\_max\_log\_gap}, is defined as
\begin{equation}
\label{eq:avg-true-max-log-gap}
\mathcal{M}_{\mathrm{gap}}
=
\frac{1}{N}\sum_{i=1}^{N}
\max\!\Bigl(
0,\;
\max_{j}\log p(j \mid i) - \log p(y_i \mid i)
\Bigr),
\end{equation}
where $N$ denotes the number of evaluated token positions, $y_i$ is the ground-truth token at position $i$, and $p(j \mid i)$ is the model-assigned probability of token $j$ at that position.

\begin{figure}[ht]
    \centering
    \includegraphics[width=\linewidth]{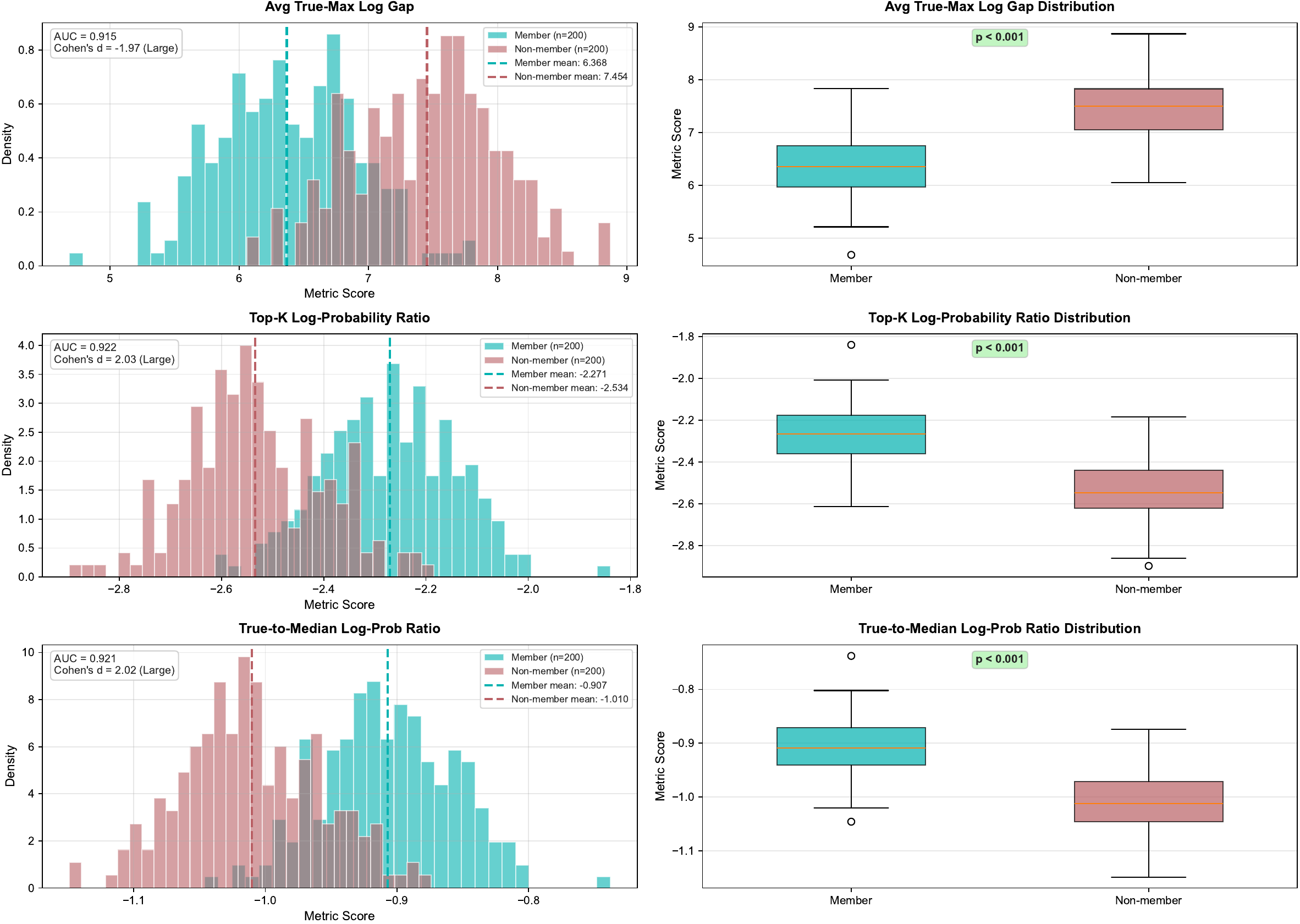}
    \caption{Validation of representative AutoMIA-discovered metrics under a controlled synthetic memorization simulation. The results show that metrics such as \texttt{avg\_true\_max\_log\_gap} produce clear separation between simulated member and non-member distributions, supporting the claim that the discovered formulas capture meaningful memorization-related structure rather than incidental correlations.}
    \label{fig:mia-metrics-validation}
\end{figure}
This metric measures the average positive log-probability gap between the model's most confident prediction and the ground-truth token. Its behavior is closely aligned with the standard intuition behind memorization. For member samples, an overfitted model is more likely to assign the highest probability to the true token, yielding
\[
\max_j \log p(j \mid i) \approx \log p(y_i \mid i),
\]
and therefore a gap close to zero. In contrast, for non-member samples, the model is less consistently aligned with the ground-truth token, which leads to a larger positive gap. Consequently, lower values of $\mathcal{M}_{\mathrm{gap}}$ correspond to stronger memorization signals.

Importantly, this quantity is not an arbitrary statistical artifact. It directly measures the extent to which the model's most confident prediction coincides with the observed target token, which is precisely the type of behavior expected when a model has memorized training examples.

\paragraph{Targeted mathematical simulation.}
To further verify that the discovered metrics respond to memorization-like structure rather than spurious correlations, we conduct a lightweight controlled simulation at the logit level.

Specifically, we construct two synthetic distributions. For the \emph{member} distribution, we inject a targeted logit boost on the ground-truth token to mimic the effect of overfitting. For the \emph{non-member} distribution, logits are sampled from a standard Gaussian distribution without such targeted reinforcement. Formally, let $\mathbf{z}_i \in \mathbb{R}^{V}$ denote the simulated logits at token position $i$ over a vocabulary of size $V$. We define
\begin{align}
\mathbf{z}_i^{(\mathrm{non})} &\sim \mathcal{N}(\mathbf{0}, I), \\
\mathbf{z}_i^{(\mathrm{mem})} &= \mathbf{z}_i^{(\mathrm{non})} + \delta \mathbf{e}_{y_i},
\end{align}
where $\mathbf{e}_{y_i}$ is the one-hot basis vector associated with the ground-truth token $y_i$, and $\delta > 0$ controls the strength of the memorization effect. We then apply the softmax function to obtain probabilities and evaluate the discovered metrics on these simulated outputs.

Under this construction, the member distribution is characterized by a stronger preference for the ground-truth token, which should reduce the value of Eq.~\eqref{eq:avg-true-max-log-gap}. This is exactly what we observe in practice. As shown in Fig.~\ref{fig:mia-metrics-validation}, \texttt{avg\_true\_max\_log\_gap} clearly separates the two synthetic distributions, assigning significantly lower scores to members, with AUC $= 0.915$, Cohen's $d = -1.97$, and $p < 0.001$. We observe similarly consistent separability for other top-ranked metrics discovered by the agent.

Taken together, these results provide complementary support from both theory and controlled simulation. They suggest that the discovered formulas are not merely fitting superficial quirks of a specific benchmark, but instead capture statistically meaningful and mechanistically interpretable signatures of memorization.

\end{document}